\documentclass[onecolumn,amsmath,amssymb,floatfix,nofootinbib]{revtex4}

\usepackage{graphicx}
\usepackage{dcolumn}
\usepackage{bm}
\usepackage{color}

\newcommand{\be}{\begin{eqnarray}}
\newcommand{\non}{\nonumber \\}
\newcommand{\ee}{\end{eqnarray}}

\begin{document}

\title{Optimal CMB estimators for bispectra from excited states}
\author{P.~Daniel Meerburg$^{1}$}
\author{Moritz M\"{u}nchmeyer$^{2,3}$}
\affiliation{$^1$CITA, University of Toronto, 60 St. George Street, Toronto, Canada}
\affiliation{$^2$Sorbonne Universit\'{e}s, UPMC Univ Paris 06, UMR7095}
\affiliation{$^3$CNRS, UMR7095, Institut d'Astrophysique de Paris, F-75014, Paris, France}

\begin{abstract}
We propose optimal estimators for bispectra from excited states. Two common properties of such bispectra are the enhancement in the collinear limit, and the prediction of oscillating features. We review the physics behind excited states and some of the choices made in the literature.  We show that the enfolded template is a good template in the collinear limit, but does poorly elsewhere, establishing a strong case for an improved estimator. Although the detailed scale dependence of the bispectra differs depending on various assumptions, generally the predicted bispectra are either effectively 1 or 2-dimensional and a simple Fourier basis suffices for accurate reconstruction. For an optimal CMB data analysis, combining all $n$-point functions, the choice for the excited state needs to be the same when computing power spectrum, bispectrum and higher order correlation functions. This has not always been the case, which could lead to wrong conclusions. We calculate the bispectrum for different  choices previously discussed for the power spectrum, setting up a consistent framework to search for evidence of excited states in the CMB data. 
\end{abstract}

\maketitle

\section{Introduction}

The mapping of the Cosmic Microwave Background (CMB) through the Planck \cite{PlanckCosmoPars2015} and WMAP \cite{WMAPfinal2012} satellites has given us an incredible insight into the evolution of the Universe. It has shown us that a simple model, counting only six parameters, can describe the observations. Two out of six of those parameters, the amplitude and first derivative of the primordial power spectrum, are necessary to describe the statistical properties of the primordial fluctuations that source the fluctuations in the CMB. The CMB power spectrum (at least in temperature) has been measured all the way up to Silk damping scales, with a precision limited only by cosmic variance. In other words, with the Planck data in combination with small scale measurements by ground based experiments such as ACT and SPT \cite{ACT2014,SPT2011,SPT2012}, we have almost reached the point were no further measure of the CMB temperature will improve our current understanding of the early Universe. Improvement can (still) be made with the measurement of polarization, both in curl free (E) and divergence free (B) modes and we expect that these will indeed lower the existing constraints on $A_s$ and $n_s$ in the coming years (as well as constraints on primordial gravitational waves). 

Theoretically, the power of the CMB lies in the fact that it potentially explores (energy) scales way beyond scales that can be reached with accelerators. It is therefore of great value and importance to squeeze every bit of information out of the CMB measurements. In particular, any deviation from a Gaussian primordial density field would provide a wealth of information on the early Universe as non-Gaussianity would for example tell us if the mechanism that produced the primordial fluctuations were driven by a weakly coupled or a strongly coupled field \cite{LSSnonGaussianity2014} or if multiple fields played a role in this process. Deviations from primordial Gaussianity have already been tightly constrained; current constraints estimate the deviations to be at most of the order of 1 in 1000 for most well motivated shape functions. The limited number of modes in the CMB prevents us from lowering current bounds on non-Gaussianity much further and we would have to use other measures, such as the large scale structure (LSS) of the Universe, to determine the nature of the field \cite{LSSnonGaussianity2014}. That being said, although profound questions about the nature of the fields will probably not be answered using CMB observations, a measurement of any $n$-point correlation function (with $n>2$) would be immensely valuable, as these measures directly probe the interactions of the field. From an EFT perspective, the most valuable bispectra are the local and equilateral shapes. However, a model of the early universe with additional degrees of freedom can easily produce other shapes. A wide class of models produces oscillating bispectra (and higher order correlation functions) that break scale invariance, besides the small spectral tilt. In a recent paper \cite{OptimalEstimator2014} we explained that it is hard to search for these bispectra as rapid oscillations result in a strong mismatch with the more common shapes, such as local and equilateral, which are only weakly scale dependent. A framework has been proposed and applied to search for any shape in the CMB temperature and polarization using a modal expansion, were any shape can be approximated using a finite number of terms on an orthogonal basis \cite{ShellardBispectrum2006,ShellardModeExpansion2009,Bean2013,JoyceShapes2015,BispectrumOscillations2010,Battefeld2011}. Although this approach is very powerful, the used 3D polynomial basis  has proven to be impractical in capturing rapid oscillations. Instead, we proposed a more case specific 1D basis which captures such bispectra with relatively few terms. As a result, we have shown that in principle we can reach frequencies for the resonant bispectrum \cite{OptimalEstimator2014} way beyond the validity of the EFT \cite{EFTOscillations2011} and at the same level as the power spectrum\cite{PlanckInflation2014,PSOscillations2011,MeerburgOscillations2014,Meerburg2014a,Meerburg2014b,PSOscillationsFlauger2013} and our estimator can be adopted to possible drifting \cite{DriftingOscillations2014} of the oscillations. 

In this paper we extend this formalism to include additional shapes. The resonant model is very specific, in the sense that its oscillations are logarithmic and depend only on the sum of the three momenta that connect the triangle ($k_t = k_1+k_2+k_3$). The literature contains a very interesting but poorly constrained class of bispectra from excited states during inflation \cite{NGFeaturesChen2007,InitialStateOriginalHolman2007,NonBDBispectrum2009,NonBDBispectrum2010,ResonantAndNonBDChen2010,NonBDBispectrum2010b,nonBDbospectrumPAgullo2011,MixedStateAgullo2011}. Unlike the power spectrum, the bispectra are a measure of the interactions. The interaction Hamiltonian can be derived from EFT arguments and induces a spectrum of shapes \cite{NGsEFT2014}. Under the assumption that the vacuum is the Bunch Davies (BD) vacuum, these are all the possible shapes. In case that the vacuum is not the BD vacuum (for example on general grounds one might expect the vacuum state to be a mixed state in a multiverse scenario \cite{MixedStateAgullo2011,Multiverse2014_Albrecht,JP2015}), the bispectrum is altered in two ways. First of all, the vacuum expectation value of the interacting fields and the interaction Hamiltonian can pick up a correction, were the correction for a pure state can be described by a Bogolyubov rotation. Secondly, the time integral of the interaction is altered as a surface needs to be specified on which the vacuum state is modified. This freedom allows for a rich phenomenology of the possible bispectra. Such bispectra from excited states typically have a momentum dependence of form $K_j = k_t-2k_j$ and include oscillations, making them an interesting target for our modified modal expansion.

This paper is organized as follows. First, in section \ref{sec:vacuumandpowerspectrum} we will review the various choices of excited states that exist in the literature. In particular, we will try to clarify some key points regarding the decay of these signatures. We then discuss the corrections to the power spectrum. In section \ref{sec:primordialbispectrum} we calculate several primordial bispectra corresponding to our state choices, and compute the resulting CMB bispectrum exactly. Based on the work in Ref.~\cite{OptimalEstimator2014,LinearOscillationsMoritz2014} we then propose a modified expansion basis to reconstruct these bispectra in section \ref{sec:modal} and compare our exact results with the expansion. We propose an optimal estimator and compute Fisher errors for one the bispectra in section \ref{sec:estimator}. We conclude in section \ref{sec:conclusion}. Throughout this paper we use a flat Planck 2013 $\Lambda$CDM cosmology with $\Omega_b h^2 =0.022$,  $\Omega_c h^2=0.12$, $h=0.68$, $\tau = 0.093$, $n_s=0.96$ and $A_s=2.2 \times 10^{-9}$.

\section{Choice of excited state and power spectrum}
\label{sec:vacuumandpowerspectrum}

\subsection{Mode equations}
\label{sec:modeequations}

The evolution of primordial curvature fluctuations $\zeta$ is governed by the Mukhanov Sasaki equation \cite{Mukhanov:1988jd,1992PhR...215..203M}
\be
\label{eq:mukasasa}
v_k'' + \left( k^2 - \frac{z''}{z} \right) v_k = 0,
\ee
with the Mukhanov variable $v(\tau)=z(\tau)\zeta(\tau)$ where $z^2 = 2 a^2 \epsilon$ and where primes indicate derivatives with respect to conformal time $\tau$. This is mathematically equivalent to an harmonic oscillator with a time dependent frequency 
\be
\omega_k^2 = k^2 - \frac{z''}{z}. 
\ee 
The scale factor $a(\tau)$ describes the isotropic homogenous background spacetime, for example for de Sitter space $a = -\frac{1}{H \tau}$ and thus $(\omega_k^{dS})^2 = k^2 - \frac{2}{\tau^2}$.

The general solution to Eq.~\eqref{eq:mukasasa} is of form 
\be
v_k = \alpha_k u_k(\tau) + \beta_k u_k^*(\tau) .
\ee
The modes must be normalized so that $W[u_k,u_k^*]=-i$ and $|a_k|^2 + |b_k|^2 = 1$. For the simple case of a de Sitter background, the positive frequency mode functions are
\be
u_k = \frac{e^{-ik\tau}}{\sqrt{2 k}}\left( 1- \frac{i}{k \tau} \right).
\label{eq:deSitterSolution}
\ee

After quantization one finds that the power spectrum of the Mukhanov variable $v$ is given by the square of the mode equations, i.e. $P_v(k,\tau) = |v_k(\tau)|^2$. The form of the positive and negative frequency modes $u_k$ and $u_k^*$ is specified by the form of the background spacetime under consideration. The complex numbers $\alpha_k,\beta_k$, called Bogolyubov parameters can be $k$-dependent. They are determined by selecting appropriate boundary conditions, and thus the state of the inflaton. We summarize choices that have been proposed in the literature.

\subsection{Review of excited states}

Before we review some of the choices found in the literature on excited states, we would like to make a general comment on what should be considered an inflationary initial state. It is sometimes argued that the effect of an excited state will redshift away. As we move further away (in time) from the excited state, corrections should be suppressed; only if modifications are relatively late w.r.t. the end of inflation should we be able to observe any effects in the spectrum of perturbations. This is true if the modified state is applied a fixed physical time $t_{\rm ini}$; indeed as it is a fixed time, not all modes will go through the same history and one introduces explicit scale dependence with any effects redshifted away from $t_{\rm ini}$. This state presents a thermal bath w.r.t. to the instantaneous vacuum and inflation should dissipate these particles as the Universe expands. In general, the longer inflation lasts, the smaller the effect; if inflation lasted longer than $\sim$70 E-folds, we expect the signal to be suppressed. { \it These excited states can be considered initial states, since they can be considered as a relic of some pre-inflationary state, such as the false vacuum state. } 

However, another possibility exists which is the consequence of considering a modification at a fixed physical scale. A mode crosses this scale at some point in time (which is different for each mode) and the mode is excited; later on it crosses another fixed physical scale, i.e. $H$, after which (for single field inflation) it is frozen. It is clear that each mode experiences the same history, besides slow-roll corrections. This is the premise of trans-Planckian physics \cite{TransPlanckianMartin2001,DanielssonTP2002,Martin:2003kp,PSOscillationsMartin2004} or the so-called new physics hyper surface (NPH) \cite{GreeneEtAl2005} scenario. We really should not refer to such a modification as an initial state modification, since in some sense we are truly adding dynamics; the physical scale excites the Universe at every moment in time during inflation. Unlike the previous case we are not exciting each mode, but only modes with $k < a(t) \Lambda$, with $\Lambda$ the scale of new physics. Instead, we should consider such a modification as an attempt to parametrize our ignorance above some physical scale, very similar to EFT. Literature suggests there is not a unique form of a rotated pure state in such a scenario, and several slightly different proposals have been considered. While each of those share the feature that they are derived from a pure state with a physical cutoff at some fixed scale, the details differ. 

In both scenarios, it is necessary to make sure that the energy of the particles does not spoil slow-roll inflation due to their gravitational back reaction on space-time. One can derive the back reaction constraints on a pure state rotation and one finds that the Bogolyubov parameter $\beta$ describing the negative frequency modes, should fall off faster than $1/k^2$ \cite{BackreactionInitialState2005,InitialStateOriginalHolman2007}. These constraints are derived when summing over all modes (UV) and one way to avoid breaking of slow-roll inflation via back reaction is to make sure no modes are excited above some cutoff scale $\Lambda$. It must be noted that such a `sharp' cutoff could potentially lead to some of the effects that we compute; if one smooths this cutoff some of the features that we compute could disappear. 

While UV modes need to fall off faster than $1/k^2$, this need not be the case for all energies. An example of this is provided by boundary effective field theory (BEFT). In this scenario the modes are affected at some fixed time (the boundary hypersurface), explicitly breaking scale invariance. However, very similar to the NPH scenario, on that boundary there exists a (high energy) physical scale. Consequently, modes close to this energy scale are affected more than IR modes. The effect is thus the opposite of what one would intuitively expect for a fixed time scenario; the modes that are most affected are at the high energy end (and thus on smaller scales in the CMB). Such a model requires a UV momentum cutoff, which may be taken to be $k_{\rm max} = a(\tau_{\rm ini}) \Lambda$. Of course, if inflation lasts long enough so that physical scales in the CMB originate above this UV cutoff, no effect of the BEFT state will be observable.

\subsubsection{Asymptotic Minkowski vacuum (Bunch Davies vacuum)}

The standard choice of vacuum is defined as the lowest energy asymptotic initial state (i.e. without particle excitations). This choice is known as the Bunch Davies (BD) vacuum. In the infinite past, $\tau \rightarrow \infty$ and every mode is deep within the horizon and much smaller than the curvature scale. One can then use the Minkowski initial conditions 
\be
\lim_{\tau \to \infty} v_k(\tau) = \frac{1}{\sqrt{2 k}} e^{-ik\tau}
\ee
which selects the positive frequency modes, i.e., $a_k =1$, $b_k=0$. In other words, the BD vacuum is the equivalent of the Minkowski vacuum in de Sitter space.

\subsubsection{Constant Bogolyubov rotation}

The algebraically simplest possible modification of the Bunch Davies vacuum is a $k$-independent Bogolyubov rotation, i.e. $a_k =c_1$, $b_k=c_2$ with constant complex coefficients so that $|a_k|^2 + |b_k|^2 = 1$. As we shall review below, such a rotation only changes the amplitude of the power spectrum, and is therefore indistinguishable from the power spectrum in Bunch Davies. However, as we shall later show, the momentum dependence of the bispectrum is modified. Physically this is because a non-zero Bogolyubov rotation implies that the modes are not in their Bunch Davies vacuum state, and therefore contain particles with respect to this state. The bispectrum measures interactions between these particles, and can therefore be enhanced.

\subsubsection{ $\alpha$-vacuum at a fixed scale and NPH scenario}

An common inflationary state proposal (see e.g. \cite{Martin:2003kp}) is to excite each Fourier mode when its wave length becomes equal to a new fundamental scale $\Lambda$ (for example the Planck scale). The model would thereby avoid to describe scales outside the range of validity of the theory. As argued earlier, such proposal is really dynamical and therefore we should not refer to it as an initial state modification. It assumes, the initial conditions for each mode are set at a $k$-dependent time $\tau_{\rm ini}(k)$ such that $k = a(\tau_{\rm ini}) \Lambda$. We set these initial conditions at a length scale and time scale so that the mode is deep within the horizon (this is possible since $\Lambda \gg H$), so that the mode equations are to good approximation of Minkowski type and we have
\be
\label{eq:bogo_minkowski}
v^{Mink}_k(\tau_{\rm ini}) = \frac{\alpha_k}{\sqrt{k}} e^{i k \tau_{{\rm ini}(k)}} + \frac{\beta_k}{\sqrt{2k}} e^{-i k \tau_{{\rm ini}(k)}}.
\ee

A natural assumption is that high energy effects should be suppressed by powers of $H/\Lambda$. The Bogolyubov parameters can then be put in the perturbative form \cite{2013CQGra..30k3001B}
\be
\label{eq:bogo_nph1}
|\alpha_k| = 1 + y \frac{H(\tau_{\rm ini}(k))}{\Lambda} + \mathcal{O}\left(\frac{H^2}{\Lambda^2}\right), \non
|\beta_k| = x \frac{H(\tau_{\rm ini}(k))}{\Lambda} + \mathcal{O}\left(\frac{H^2}{\Lambda^2}\right).
\ee

Our guiding principle here is to construct initial conditions that are exactly scale invariant in de Sitter space. Therefore we assumed that the magnitude of the Bogolyubov parameters depends on $k$ only implicitly due to $H$ (when breaking de Sitter invariance by making $H$ a function of $k$), and $x$ and $y$ are k-independent real constants, naturally of order $1$. In addition, the relative phase of the positive and negative frequency solution must be chosen scale invariantly, since it is not an overall phase of the quantum mechanical state and therefore observable in the power spectrum, as we shall review in the next section. In de Sitter space we have $\tau_{\rm ini} = -\frac{\Lambda}{kH}$.  The unique scale invariant vacuum choice at this order in $H/\Lambda$ is therefore given by the Bogolyubov parameters 
\be
\label{eq:bogo_nph2}
\alpha_k = \left( 1 + y \frac{H}{\Lambda} \right) e^{i \frac{\Lambda}{H}}  \hspace{1cm} \beta_k = \left( x \frac{H}{\Lambda} \right) e^{-i (\frac{\Lambda}{H}+\phi)} .
\ee
Here we chose the phases of the modes so that the $\alpha$-modes start at $\tau_{ini}$ in phase $0$ and the $\beta$-modes start in the $k$-independent phase $\phi$ (when inserting Eq.~\eqref{eq:bogo_nph2} in Eq.~\eqref{eq:bogo_minkowski}).

To compare our result with that of Ref.~\cite{GreeneEtAl2005}, we rewrite the mode equation in the form $v = N(k) (u_k + b(k) u_k^*)$ and obtain
\be
b(k) = x \frac{H(\tau_{\rm ini}(k))}{\Lambda} e^{-i \left(2 \frac{\Lambda}{H(\tau_{\rm ini}(k))}+\phi \right)}
\ee
in agreement with the Bogolyubov parameter of the new physics hyper surface (NPH) scenario. They also note that in this model the exponential factor appears to be necessary to avoid non-localities to order $H$ (while here it arises due to scale invariance). Both the NPH scenario and the trans-Planckian argument are based on the same principle; there is some physical scale (NPH: $\Lambda$, trans-Planckian: $M_p$) beyond which we assume each mode rotates off the Bunch Davies vacuum, i.e. is excited. Since this is a fixed scale, very similar to the Hubble scale during inflation, every mode will experience the same history, up to slow-roll corrections due to the broken de Sitter symmetry. Any mode that exited the horizon before set scale is not affected and is assumed to be in the vacuum state; however these modes are unobservable as they are beyond our current horizon.

\subsubsection{Fixed time vacua and the BEFT scenario}

A true initial state modification sets the initial conditions at a $k$-independent time $\tau_{\rm ini}$, for example the earliest time in which we can trust general relativity. The initial conditions are then explicit functions of $k$, breaking scale invariance. In this scenario, corrections to the Minkowski vacuum should come as powers of of the physical wave number $k^{phys}_{\tau_{\rm ini}}=k/a_{\rm ini}$ divided by a cutoff scale $\Lambda$. The Bogolyubov parameters are then
\be
|\alpha_k| = 1 + y  \frac{k}{a_{\rm ini}\Lambda}  + \mathcal{O}\left( \frac{k}{a_{\rm ini}\Lambda}\right), \non
|\beta_k| = x \frac{k}{a_{\rm ini}\Lambda} + \mathcal{O}\left( \frac{k}{a_{\rm ini}\Lambda} \right).
\ee
We assume that all modes start at $\tau_{\rm ini}$ at a fixed k-independent phase. Therefore we find
\be
\alpha_k = \left( 1 + y \frac{k}{a_{\rm ini}\Lambda} \right) e^{-i k \tau_{\rm ini}}  \hspace{1cm} \beta_k = \left( x \frac{k}{a_{\rm ini}\Lambda} \right) e^{i (k \tau_{\rm ini}+\phi)}.
\ee
Unlike in the NPH case, oscillations in the power spectrum here also arise in pure de Sitter space, where $H$ is constant. The Bogolyubov rotation in so-called boundary effective field theory (BEFT) \cite{BEFTInitialState2005} was first parametrized in Ref.~\cite{GreeneEtAl2005} 
\be
b(k) = x \frac{k}{a_{\rm ini}\Lambda} e^{-i \left( 2\frac{k}{H a_{\rm ini}} +\phi\right)}.
\ee
Compared with the previous case, the non-BD vacuum here explicitly breaks scale dependence. Since there is both a fixed time and a fixed scale, different modes, frozen when they cross the horizon, have not experienced the same history. As we discussed above, the BEFT scenario can generate large effects on UV scales, which is a result of the assumption that on the initial hypersurface, modes that have a momentum closer to the set UV cutoff scale, will be more affected. Because of the form of the Bogolyubov rotation, we expect this modes to be best constraints using smaller scales, for example those probed by LSS at high redshift. However, given the back reaction constraints, modes $k>k^{\rm pays}_{\rm max}$ at some time $a_{\rm ini}$ are not excited. Although not explicit in this equation, we should keep in mind, if this is too early, the comoving cutoff will eventually redshift outside of our current horizon and we will not be able to observe these effects at all.

\subsection{Shape of the primordial power spectrum}

As discussed by Ref.~\cite{GreeneEtAl2005} the Bogolyubov parameter $b(k)=|b(k)|e^{\alpha(k)}$ defined in $v_k = N(k) (u_k + b(k) u_k^*)$  immediately leads to a power spectrum of the curvature perturbations of form 
\be
P(k) \simeq P_{BD}\left(1+2 |b(k)| \cos(\alpha(k) + \psi)\right).
\ee
A constant Bogolyubov rotation thus only results in a renormalization of the power spectrum (at lowest order in slow roll). For the NPH initial conditions discussed before we get
\be
P^{NPH}(k) \simeq P_{BD}\left(1+2 x \frac{H(\tau_{\rm ini}(k))}{\Lambda}  \cos(\frac{\Lambda}{H(\tau_{\rm ini}(k))} + \psi)\right),
\ee
where $H(\tau_{\rm ini}(k)) \propto \frac{H_0}{\epsilon \log (k/k_0)}$. We thus obtain  logarithmic oscillations in $k$.
The BEFT scenario gives
\be
P^{BEFT}(k) \simeq P_{BD}\left(1+2 x \frac{k}{a_{\rm ini}\Lambda}  \cos(\frac{k}{H_{\rm ini} a_{\rm ini}} + \psi)\right),
\ee
where $H_{\rm ini}$ is constant at $\tau_{\rm ini}$, so we obtain linear oscillations in $k$. The phase $\psi$ is arbitrary and $k$-independent. 

Generally, if the solution to the e.o.m. for the inflation degree of freedom can be written as a super position of positive and negative frequency modes, and if there is scale dependent phase, oscillations in the power spectrum will appear \cite{GreeneEtAl2005,MonodromyFlauger2009,OptimalEstimator2014}. Consequently, if the breaking of scale invariance is at most slow-roll, models are hard to distinguish at the level of the power spectrum. Higher order correlation spectra proof a way out, since they are sensitive to the dynamics of the field and as such, allows us to discriminate between models.

\section{Shape of the primordial bispectrum}
\label{sec:primordialbispectrum}

\subsection{General remarks for arbitrary $H_{int}$}

\subsubsection{Leading order bispectrum}

The bispectrum at tree level for a Gaussian initial state is given by \cite{Maldacena:2002vr}
\be
\langle \zeta_{{\bf k}_1}\zeta_{{\bf k}_2}\zeta_{{\bf k}_3} \rangle &=& -i \int d\tau \langle \left[ \zeta_{{\bf k}_1}\zeta_{{\bf k}_2}\zeta_{{\bf k}_3},H_{\rm I}(\tau)\right]\rangle  \nonumber \\
&=& -2 \mathcal{R}e \int d\tau i \langle \zeta_{{\bf k}_1}\zeta_{{\bf k}_2}\zeta_{{\bf k}_3} H_{\rm I}(\tau)\rangle, 
\ee
with $H_{\rm I}$ the interaction Hamiltonian. The Hamiltonian is theory specific as are the solution to the equations of motion for $\zeta$ (which by itself are derived from the varying the action w.r.t. to $\zeta$). We start with general remarks about this calculation, before evaluating the bispectrum for the simplest cases.

Without further details on $H_{\rm I}$ we can consider solutions for $\zeta$ that are rotated w.r.t. to the usual Bunch Davies vacuum, i.e. are in an excited state. Each rotated $\zeta$, will now contain some particles, and the density is determined by the Bogolyubov parameter $\beta$. Unlike the power spectrum, a constant $\beta$ {\it already} changes the momentum dependence of the bispectrum (while for the power spectrum a $k$-independent rotation would lead to a renormalization of the power spectrum). This is sourced by the fact that interactions take place during inflation and the strength of this effect is determined by the interaction Hamiltonian as well as the total time that inflation lasts. More precisely, the presence of particles and the interaction affect both the shape and amplitude of the primordial curvature 3-point function.

If the solutions of the e.o.m. for $\zeta$ are a superposition of positive and negative frequency modes as explained in section \ref{sec:modeequations}, then given the Hamiltonian to cubic order in $\zeta$, and 3 $\zeta$'s for the 3 vertexes we then generally find that the 3 point function can be written as (to lowest order in the $\beta$ and with $\alpha \sim 1$)
\be
\label{eq:leadingnonbd}
\langle \zeta_{{\bf k}_1}\zeta_{{\bf k}_2}\zeta_{{\bf k}_3} \rangle &\sim& \mathcal{O}(\beta^0)+ c_1(k)\delta(\sum {\bf k}_i) \int d\tau \left[a^n \sum_j\beta_{k_j}^* e^{i K_j \tau} + {\rm c.c.} \right]+ \mathcal{O}(\beta^2),
\ee
with $K_j \equiv k_t -2k_j$ and the amplitude of $c_1$ depends on the strength of the interaction and its scale dependence is such that for the chosen interaction (and thus the power $a^n$) the overall scaling of the bispectrum is as $1/k^6$\footnote{If the cutoff time is scale independent, one explicitly breaks scale invariance and this argument no longer holds.}. Note that since this is a connected 3 point function, that the only surviving contraction are those with 3 distinct (ingoing) momenta, the other 3 momenta serve as conservation of total momentum. 

Note that for $n=0$ we have $ c_1(k)\propto 1/k^5$. A more complicated interaction Hamiltonian will generally lead to different power of $a$, and hence different pre-factors in powers of $K_j$. This is crucial, as different powers of $a$ also dictate the maximal enhancement \cite{InitialStateOriginalHolman2007,NonBDBispectrum2009}; for $n$ different from 0 one has to partially integrate to perform this integral and therefore higher power of $K_j$ will appear in the denominator. The higher the power the more powers it can `cancel' in the numerator, e.g. 
\be
\lim_{K_j \rightarrow 0} \frac{\sin \omega K_j}{K_j} = \omega,
\ee
while 
\be
\lim_{K_j \rightarrow 0} \frac{1-\cos \omega K_j}{K_j^2} = \omega^2/2,
\ee
etc. Since $\omega \gg 1$ the enhancement in the collinear limit is dictated by the power of $a$.

\subsubsection{Choice of initial time surface and $k$ dependence of the Bogolyubov parameter}

To evaluate the shape via Eq.~\eqref{eq:leadingnonbd}), there are two choices to be made that were not present in the power spectrum calculation, and that influence the resulting shape. These are the initial time surface $\tau_{\rm ini}(k)$ and the momentum dependence of the Bogolyubov parameter (which can now depend on either $k_i$ or $k_t$ for example). 

We first discuss the choice of initial time. It is clear that the above integral diverges for finite momenta if we take the conformal time integral all the way to negative infinity (except for the part inside $\mathcal{O}(\beta^0)$). From back reaction constraints it was obtained \cite{BackreactionInitialState2005,InitialStateOriginalHolman2007} that $\beta$ should fall off faster than $1/k^2$. Here that is achieved by making sure that no modes are excited for $k>a\Lambda$ Unlike the power spectrum, we now have to be specific about our initial time. We shall see that the choice here is crucial and determines the scale dependence of the bispectrum. We consider several possible choices, in line with our discussion of the effects on the power spectrum. The most obvious choice for the initial time is to place a finite time cutoff $\tau_0$, independent of the momentum $k$. However, the following argument suggests another, momentum dependent initial time surface. For all positive frequency modes, we should choose an initial condition that returns the bispectrum of single field slow-roll model. The contribution from the positive frequency modes goes like
\be
\label{eq:inin}
\langle \zeta_{{\bf k}_1}\zeta_{{\bf k}_2}\zeta_{{\bf k}_3} \rangle_{\beta^0} \sim c_2(k)\delta(\sum {\bf k}_i) \int d\tau \left[a^n e^{i k_t \tau} + {\rm c.c.} \right].
\ee 
For the Bunch Davies choice, one typically gets terms that go like $1/k_t$. However, when we apply a modification at some time $\tau_{\rm ini}$ we get additional scale dependence introduced by the cutoff; these corrections are of the form 
\be
\frac{\sin k_t \tau_{\rm ini}}{k_t},
\ee
and a possible choice of $\tau_{\rm ini}$ could be $n\pi/k_t \times \Lambda/H$, which makes sure this term vanishes and all positive frequency modes lead to the solution associated with BD initial conditions. One can argue for a third choice. As we have seen above, once we add negative frequency solutions, we introduce a problem since an excited vacuum would give infinite contribution to the 3-point function at early times. From this perspective it is the negative frequency mode that should determine the initial time surface, so one should choose a cutoff that renders this direction a `special' direction and $\tau_{\rm i}= 1/k_1 \times \Lambda/H$. This is slightly different from the choice made earlier ($\tau_{\rm ini}(k_t)$) and would generate an excited state earlier since $k_j<k_t$, so its effect should be bigger. We would like to point out that with all three choices, $\tau_{\rm ini} = {\rm const.}$, $\tau_{\rm ini}(k_t)$, $\tau_{\rm ini}(k_i)$, the computation relies on a sharp cutoff. Such a cutoff might not be completely natural. In case the boundary conditions are softer, one would expect that some of the  features computed here  might be damped.

The momentum dependence of the Bogolyubov parameter is also crucial. Even when using a constant $k$-independent Bogolyubov parameter, one gets oscillations in the bispectrum. This is in contrast to the situation in the power spectrum. For example, the characteristic logarithmic oscillations in the NPH power spectrum explicitly require $H=H(k)$, i.e. $H$ is broken by slow-roll and the associated Bogolyubov parameter obtains a scale dependent phase. In previous works it was generally assumed that the Bogolyubov rotation was constant and hence did not lead to additional scale dependence; eventually, one would like to compare the observations of the power spectrum that of higher order correlations functions (e.g. the bispectrum) and for that reason we will also consider a Bogolyubov parameter that is $k$-dependent, consistent with the choices made for the power spectrum. If the Bogolyubov parameter is explicitly momentum dependent, as is the case in BEFT, the natural momentum dependence is $\beta(k_i)$, where $k_i$ is the momentum of the corresponding leg in the excited state. If on the other hand $\beta$ depends on the momentum only implicitly by $H=H(k)$, as in the NPH scenario, we use the total momentum of the three-point interaction, i.e. $\beta(H(k_t))$. The time cutoff $\tau_{\rm ini}$ in the NPH scenario is naturally chosen with the same momentum dependence. To summarize, for the NPH scenario we will use $\tau_{\rm ini}(k_t)$ and $\beta(H(k_t))$, and for the fixed time BEFT \cite{BEFTInitialState2005} scenario we use  $\tau_{\rm ini}=const.$ and $\beta(k_i)$. This choice also guarantees that the resulting shapes are at most effectively two-dimensional in their modal expansion to be proposed below, independent of the powers of $k$ coming from the interaction Hamiltonian.

\subsection{Single field slow-roll non Bunch Davies shapes}

We now specialize our above discussion to the single field slow-roll interaction Hamiltonian \cite{Maldacena:2002vr}
\be
H_I=-\int d^3 x\ a^3\ \left(\frac{\dot{\phi}}{H}\right)^4 H {\zeta'}^2 \partial^{-2}\zeta',
\ee
which captures the dominant non-Gaussianity contribution in a single vertex. While single field slow-roll is well known to give small non-Gaussianities, we nevertheless get insights into the possible oscillating $k$-dependencies induced by non-Bunch Davies vacua.  We are using de Sitter mode functions, as in the case of the power spectrum. While corrections to the mode functions arise when $\dot{H} \neq 0$, as in the case of the NPH scenario, such corrections are slow-roll suppressed. From Eq.~ \eqref{eq:inin}, to linear order in $\beta_k$, the bispectrum correction due to the non-Bunch Davies vacuum is for this interaction Hamiltonian \cite{InitialStateOriginalHolman2007,NonBDBispectrum2009}
\be
 \Delta \langle \zeta_{\vec{k}_1} \zeta_{\vec{k}_2}\zeta_{\vec{k}_3} \rangle = 
-i (2\pi)^3 \delta^3(\sum \vec{k}_i) \frac{2}{\prod (2k_i^3)} \frac{H^6}{M_{\rm pl}^2\dot{\phi}^2}  \int^0_{\tau_{\rm ini}} d\tau \sum_j \beta^*_{k_j}
\frac{k_1^2k_2^2k_3^2}{k_j^2} e^{i K_j \tau} +{\rm c.c.}. 
\ee 
As reviewed above, there are different possible choices for the $k$-dependence of $\tau_{\rm ini}$ and $\beta_k$, which we will discuss next.

\subsubsection{Constant Bogolyubov parameter, constant or $k$-dependent $t_{\rm ini}$}

The algebraically simplest result is obtained for a constant Bogolyubov parameter $\beta=e^{i \phi}$ and a constant initial time $\tau_{\rm ini}$. In this case, the integral evaluates to 

\be 
\label{eq:shape_nonbd1}
\Delta \langle \zeta_{\vec{k}_1} \zeta_{\vec{k}_2}\zeta_{\vec{k}_3} \rangle_{\mathrm{nBD1}} = 
-(2\pi)^3 \delta^3(\sum \vec{k}_i)  \frac{4 H^6}{M_{\rm pl}^2\dot{\phi}^2}
 \frac{k_1^2k_2^2k_3^2}{\prod (2k_i^3)} \sum_j \frac{ \cos(\phi)-\cos( \omega K_{j}+\phi) } {k_j^2 K_j},
\ee
where the oscillation frequency is formally given by the initial time, $\omega=\tau_0$. In the case of $\tau_{\rm ini}={\rm const.}$, in principle also the term with three legs in the BD state is modified, and acquires an oscillation in $\cos(\omega k_t)$. This form of oscillatory shape was discussed in Ref.~\cite{LinearOscillationsMoritz2014} where the CMB multipole bispectrum was calculated and an optimal estimator was proposed. 

If the initial time is chosen as $\tau_{\rm ini}\propto \frac{1}{k_t}$, the shape is modified to be
\be 
\label{eq:shape_nonbd2}
\Delta \langle \zeta_{\vec{k}_1} \zeta_{\vec{k}_2}\zeta_{\vec{k}_3} \rangle_{\mathrm{nBD2}} = 
-(2\pi)^3 \delta^3(\sum \vec{k}_i)  \frac{4 H^6}{M_{\rm pl}^2\dot{\phi}^2}
 \frac{k_1^2k_2^2k_3^2}{\prod (2k_i^3)} \sum_j \frac{ \cos(\phi)-\cos(( \omega \frac{K_{j}}{k_t})+\phi) } {k_j^2 K_j},
\ee
which changes the directional dependence of the oscillation in $k$-space.

\subsubsection{NPH scenario}

Now we assume an NPH vacuum, i.e. $\beta_{k} = \left( x \frac{H}{\Lambda} \right) e^{-i \frac{\Lambda}{H}} $ (dropping a possible constant phase $\phi$ for notational simplicity) and $\tau_{\rm ini} = \Lambda/(H k_t)$. We find 
\be 
\Delta \langle \zeta_{\vec{k}_1} \zeta_{\vec{k}_2}\zeta_{\vec{k}_3} \rangle_{\mathrm{nBD3}} = 
-(2\pi)^3 \delta^3(\sum \vec{k}_i)  \frac{4 H^6}{M_{\rm pl}^2\dot{\phi}^2}
 \frac{k_1^2k_2^2k_3^2}{\prod (2k_i^3)} \sum_j \left( x \frac{H}{\Lambda} \right) \frac{ \cos(\frac{\Lambda}{H}) -\cos( \frac{\Lambda}{H} (\frac{K_{j}}{k_t} + 1)) } {k_j^2 K_j}
\ee,
which has the correct limiting behavior for $K_j\rightarrow 0$, i.e.,
\be
\lim _{K_j\rightarrow 0}\Delta \langle \zeta_{{\bf k}_1}\zeta_{{\bf k}_2}\zeta_{{\bf k}_3} \rangle  \propto \delta^3(\sum {\bf k}_i)\frac{1}{k_1k_2k_3} \sum_j \frac{\sin( \Lambda/H)}{k_j^3}.
\ee

So far we have not taken into account the breaking of de Sitter symmetry in the background, that is necessary to obtain the logarithmic oscillations in the NPH power spectrum. In the case of the power spectrum, this could be taken into account by making $\beta$ implicitly depending on $k$ via $H$. To be completely consistent, one should calculate the NPH bispectrum in the slow-roll expansion, to take into account both the slow-roll corrections to the mode functions and the slow-roll corrections in the vacuum rotation $\beta_{k_t}$. Here we only take into account the latter of these effects and obtain

\be 
\label{eq:shape_nonbd4} 
\Delta \langle \zeta_{\vec{k}_1} \zeta_{\vec{k}_2}\zeta_{\vec{k}_3} \rangle_{\mathrm{nBD4}} = 
-(2\pi)^3 \delta^3(\sum \vec{k}_i)  \frac{4 H^6}{M_{\rm pl}^2\dot{\phi}^2}
 \frac{k_1^2k_2^2k_3^2}{\prod (2k_i^3)} \sum_j \left( x \frac{H_\star}{\Lambda} \right) \frac{ \cos(\frac{\Lambda}{H_\star} \log k_t) -\cos( \frac{\Lambda}{H_\star} (\frac{K_{j}}{k_t} + 1) \log k_t) } {k_j^2  K_j \log k_t}.
\ee

In the collinear limit we then obtain
\be
\lim _{K_j\rightarrow 0}\Delta \langle \zeta_{{\bf k}_1}\zeta_{{\bf k}_2}\zeta_{{\bf k}_3} \rangle &\propto& \delta(\sum {\bf k}_i) \frac{1}{k_1k_2k_3} \sum_j \frac{\sin (\Lambda/H_\star \log k_t)}{k_j^3},
\ee
which resembles resonance non-Gaussianities.

\subsubsection{Fixed time vacua with $\beta_k$}

In this case we explicitly break scale invariance in the initial conditions by considering $\beta_k$ as in the BEFT scenario \cite{BEFTInitialState2005}. For a fixed time vacuum at  $\tau_{\rm ini}$ with Bogolyubov parameter $\beta_{k_j} = \left(\frac{x k_j}{a_{\rm ini}\Lambda} \right) e^{i k_j \tau_{\rm ini}}$ we get linear oscillations of form
\be 
\Delta \langle \zeta_{\vec{k}_1} \zeta_{\vec{k}_2}\zeta_{\vec{k}_3} \rangle_{\mathrm{nBD5}} = 
-(2\pi)^3 \delta^3(\sum \vec{k}_i)  \frac{4 H^6}{M_{\rm pl}^2\dot{\phi}^2}
 \frac{k_1^2k_2^2k_3^2}{\prod (2k_i^3)} \sum_j \left( x \frac{k_j}{a_{\rm ini} \Lambda} \right) \frac{ \cos \left(\tau_{\rm ini} k_j\right) -\cos\left( \tau_{\rm ini} (k_j + K_j) \right) } {k_j^2 K_j}.
\ee
As is the case in the power spectrum, we find that for fixed time initial states the non-BD contribution of the bispectrum is more suppressed the smaller the physical momentum $k/a_{\rm ini}$ is compared to the energy scale $\Lambda$ (and therefore the larger the scale on the CMB sky). Note that in principle there could be additional contributions from the boundary itself as discussed in \cite{BoundaryNGsPorrati2004}.

\subsection{Exact calculation of example CMB multipole bispectra}
\label{sec:exactcalc}

\begin{figure}
\resizebox{0.9\hsize}{!}{
\includegraphics{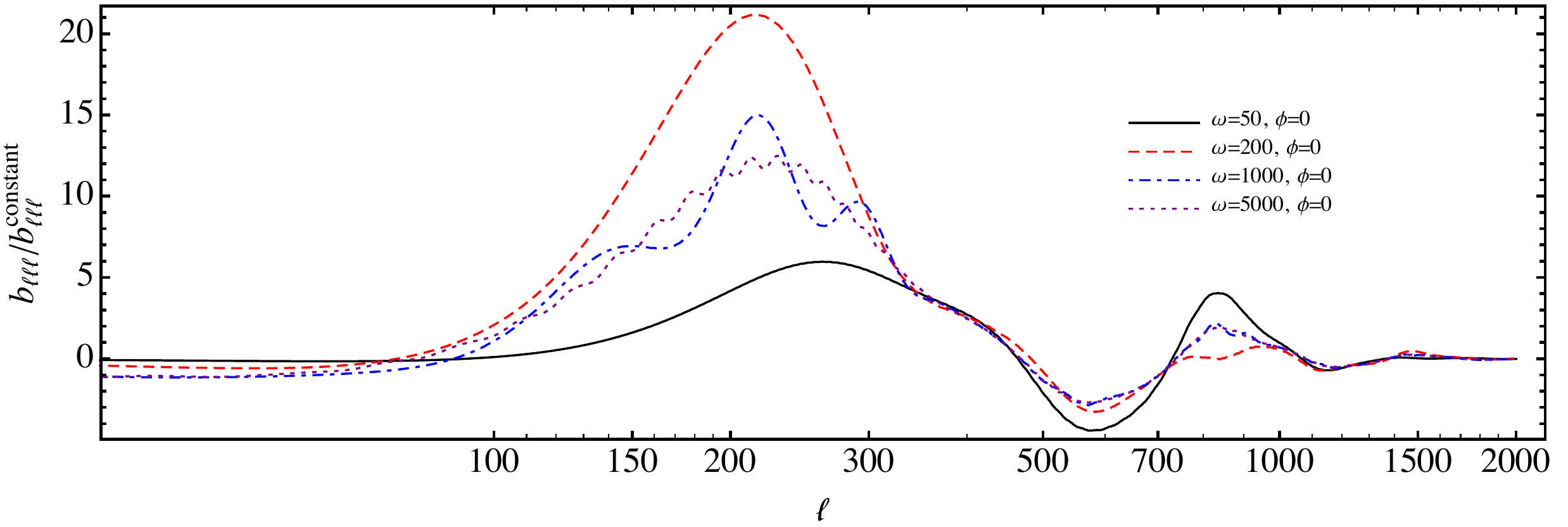}
}
\caption{The shape of Eq.~\eqref{eq:simplebd} on the diagonal $\ell_1=\ell_2=\ell_3$ for 4 different frequencies. The constant determines the overall scaling with $\ell$, and superimposed oscillations appear on top of the BAO.  }
\label{fig:nonBD_shape1_equilateral_lmax2000}
\end{figure}

\begin{figure}
\resizebox{0.92\hsize}{!}{
\includegraphics{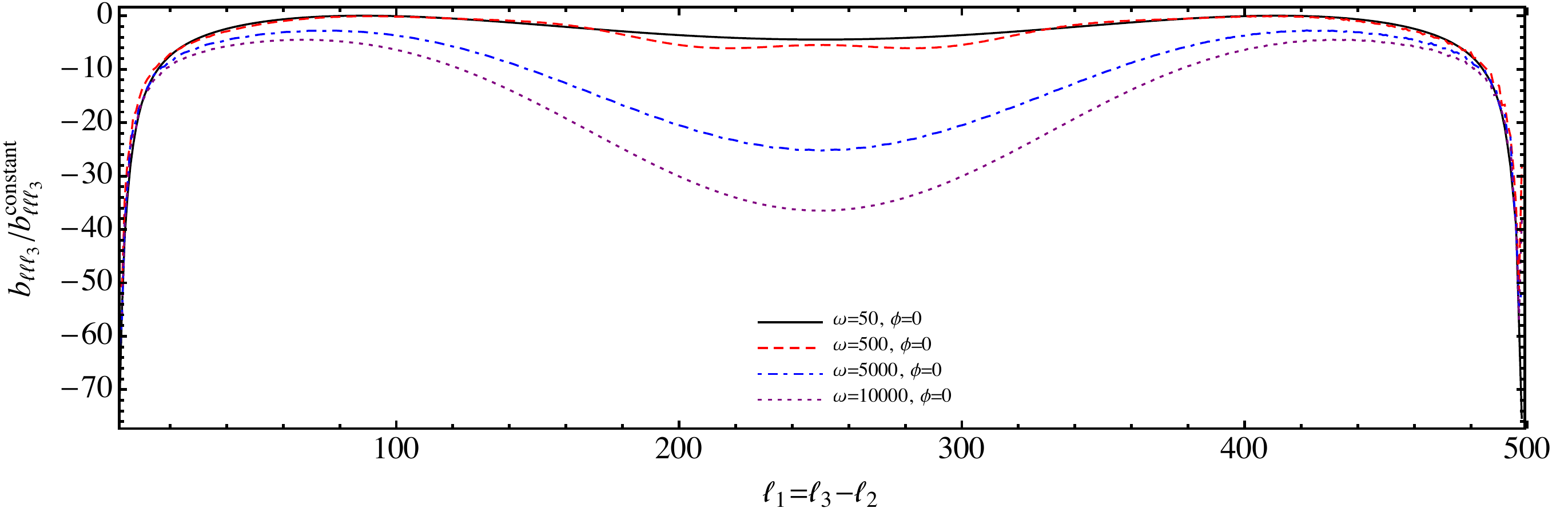}
}
\caption{The shape of Eq.~\eqref{eq:simplebd} in the collinear limit $\ell_3 = 500 = \ell_1+\ell_2$. Unlike the diagonal there are no visible oscillations, even when you increase the frequency. }
\label{fig:nonBD_shape1_collinear_lmax500}
\end{figure}

In this section we evaluate exact CMB multipole bispectra for some of the shapes discussed in the previous section. We use these exact calculations to obtain an intuition about what can be seen in the CMB as well as to verify our modal expansion proposed in the next section.

The simplest case is the shape derived in Eq.~\eqref{eq:shape_nonbd1} with corresponding normalized bispectrum (assuming here phase $\phi=0$)
\be
\label{eq:simplebd}
B(k_1,k_2,k_3) = \frac{k_1^2k_2^2k_3^2}{k_1^3k_2^3k_3^3} \sum_j\frac{1-\cos \omega K_j}{k_j^2K_j}.
\ee

We can use the original re-parametrization proposed by Fergusson and Shellard \cite{ShellardBispectrum2006}:
\be
k_1 &=& ka = k(1-\beta), \nonumber \\
k_2 &=& kb = \frac{1}{2}k(1+\alpha+\beta), \nonumber  \\
k_3 &=& kc =  \frac{1}{2}k(1-\alpha+\beta). 
\ee 
Note that $k_1 + k_2 + k_3 = k_t = (a+b+c) k = 2k$, hence $k = k_t/2$. The parameters have the following domains $0 \leq k \leq \infty$, $ 0 \leq \beta \leq 1$ and $-(1-\beta) \leq \alpha \leq (1-\beta)$. The volume element can be computed through the determinant of the Jacobean, i.e. ${\rm Det}\;J_{ij} = k^2$, and $dk_1dk_2dk_3 = k^2 dk d \alpha d\beta$. 
Unlike the resonant shape, the example shape above now explicitly introduces an non-factorizable form in the $\alpha$-$\beta$ plane. This is not a real problem, since the original integrals over the Bessel and transfer function already require sampling both in $\alpha$ and in $\beta$ in one loop.  One can actually propose a very similar parameterization to avoid this, but because there is really no computational advantage, we decided to stick to the original parameterization. It is not hard to show that the shape is then given by 
\be
S(q,\alpha,\beta) &=& \frac{1}{{a}^3 b c}\frac{(1-\cos 2 q (1-a) \omega)}{(1-a)} + \rm{perm}.  
\ee
In the shape above, $a$ can be 1 and numerically this can lead to some issues which are easily resolved by explicitly writing down the $a,b,c\rightarrow 1$ limit. As it should, that limit is finite and does not diverge; It is enhanced in the squeezed limit but not beyond local \cite{ExcitedLimitBispectra}. 

In Fig.~\ref{fig:nonBD_shape1_equilateral_lmax2000} we show the projected shape  one the equilateral axis $\ell_1=\ell_2=\ell_3$ divided by the constant bispectrum 
\be
b_{\ell_{1}\ell_{2}\ell_{3}}^{\rm constant} & = &\frac{1}{(2\ell_{1}+1)}\frac{1}{(2\ell_{2}+1)}\frac{1}{(2\ell_{3}+1)}\left[ \frac{1}{\ell_{t}+3}+\frac{1}{\ell_{t}}\right]
\ee
From the figure. it appears the shape is dominated by the constant, which simply leads to BAOs in the bispectrum. As we increase the frequency of the oscillations, the shape roughly shows the same BAO structure, with superimposed oscillations, very similar to what happens to the power spectrum. Also note that for $\omega = 50$ partial cancellation seems to occur between the oscillating part and the BAO, resulting in overall lower amplitude. 

In Fig.~\ref{fig:nonBD_shape1_collinear_lmax500} we show the collinear limit, with $\ell_3 = 500$ and $\ell_1+\ell_2=\ell_3$. This is the same limit in which the primordial spectrum maximizes, i.e. corresponding to $K_j\rightarrow 0$. Again we should not be concerned with absolute value of the amplitude, as it is arbitrary. Unlike the equilateral limit there are no clear visible oscillations on top of a smooth spectrum. Even when you increase the the frequency, no clear features appear, suggesting that the collinear limit is indeed dominated by a constant, i.e. $\omega$. In the collinear limit, the peaks appear when either one out of 3 $\ell$ gets large, i.e. $\ell_2 \ll \ell_3 \sim \ell_1$ (and symmetric equivalent) and as the frequency is increased $\ell_2\sim \ell_1 = \ell_3/2$. 

\begin{figure}
\resizebox{0.9\hsize}{!}{
\includegraphics{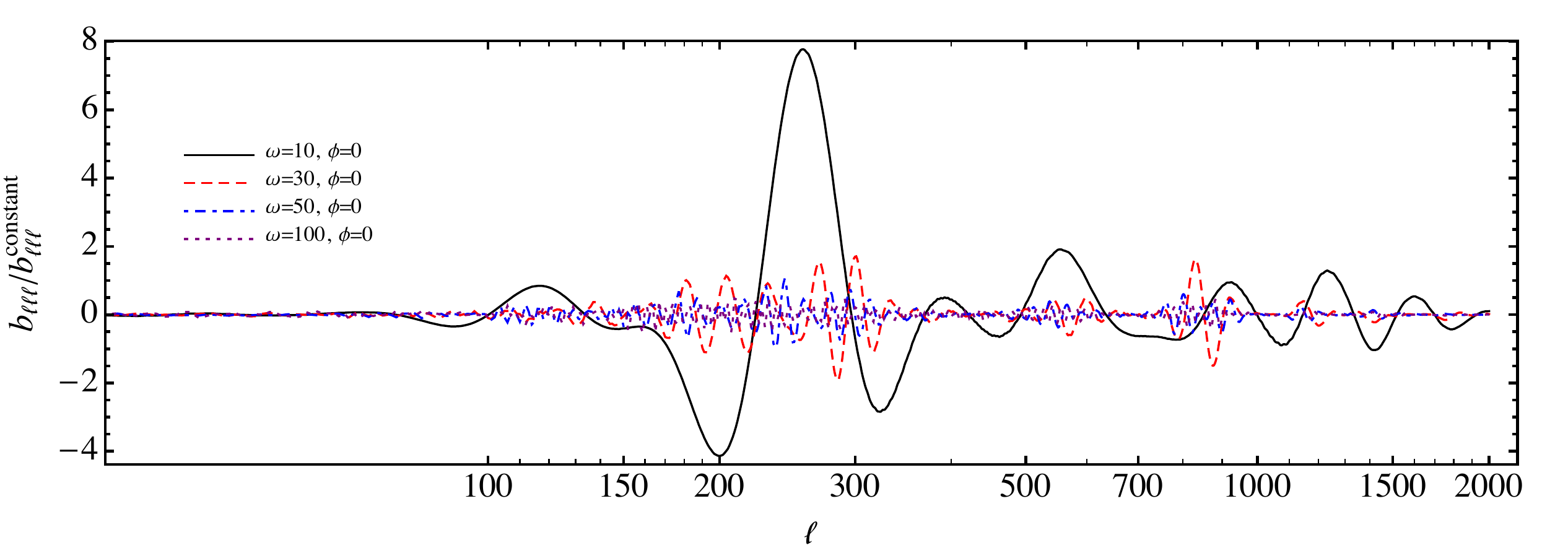}
}
\caption{The shape of Eq.~\eqref{eq:lesssimplebd} on the diagonal $\ell_1=\ell_2=\ell_3$ for 4 different frequencies. Compared to the shape in Fig.~\ref{fig:nonBD_shape1_equilateral_lmax2000} the constant has now been replaced with a cosine oscillating in $\log k_t$; this leads to a shape that resembles something like the resonant bispectrum in the equilateral limit.}
\label{fig:nonBD_shape2_equilateral_lmax2000}
\end{figure}

We will consider another example, in case the Bogolyubov rotation and the initial time $\tau_{\rm ini}$ both depend on $k_t$ as in Eq.~\eqref{eq:shape_nonbd4} we have\footnote{Inside the log, $k_t$ should be dimensionless, so we choose some pivot scale $k_* = 1$ Mpc$^{-1}$. }
\be
\label{eq:lesssimplebd}
B(k_1,k_2,k_3) = \frac{k_1^2k_2^2k_3^2}{k_1^3k_2^3k_3^3} \frac{1}{k_t} \sum_j\frac{\cos \omega \log k_t-\cos \omega (1-\frac{2k_j}{k_t})\log k_t}{k_j^2(1-\frac{2k_j}{k_t})\log k_t}
\ee
We find
\be
S(q,\alpha,\beta) &=& \frac{1}{a^3bc}\frac{\cos \omega \log 2q-\cos  (\omega (1-2a)\log 2q)}{(1-a)\log 2 q} + \rm{perm.},
\ee
with the limiting behavior
\be
\lim_{a,b,c\rightarrow 0} S(q,\alpha,\beta) =  \frac{1}{a^3bc} \omega \sin(\omega \log 2q)+ \rm{perm.}
\ee
We show the result for several frequencies in the equilateral limit in Fig.~\ref{fig:nonBD_shape2_equilateral_lmax2000}. The dominant contribution is now set by the cosine oscillating in $\log k_t$. It therefore resembles the resonant bispectrum in this limit, however the shape is modified since the smooth scale dependence now is a function of $K_j$. 

\begin{figure}
\resizebox{0.92\hsize}{!}{
\includegraphics{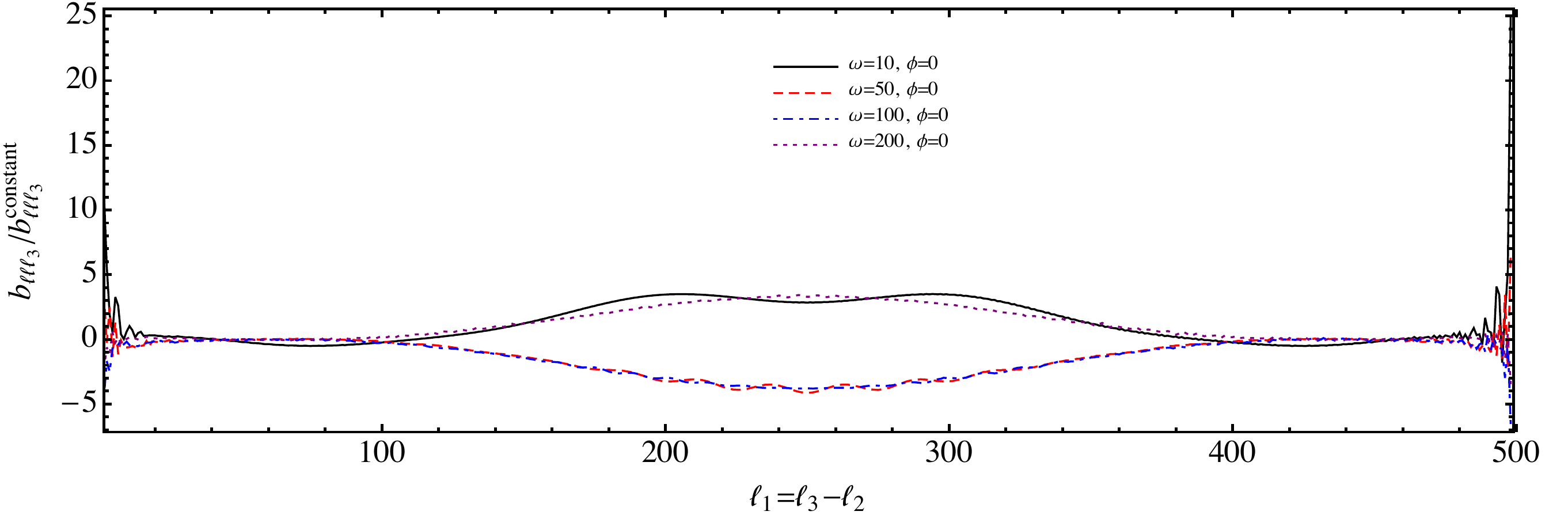}
}
\caption{The shape of Eq.~\eqref{eq:lesssimplebd} in the collinear limit $\ell_3 = 500 = \ell_1+\ell_2$. Small features appear on all scales, with strong features in the squeezed limit. }
\label{fig:nonBD_shape2_collinear_lmax500}
\end{figure}

In Fig. \ref{fig:nonBD_shape2_collinear_lmax500} we show the collinear limit for $\ell_3 =\ell_1+\ell_2 = 500$. For the frequencies considered the bispectrum changes sign, but hardly changes amplitude; it does however appear to contain rapid oscillations in the squeezed limit. 

\section{Mode decomposition and estimator for the CMB bispectrum}
\label{sec:modal}

\subsection{The case for modal expansion}

A template shape, called the enfolded shape, has been proposed to recover the collinear behavior of the shapes above \cite{NonBDBispectrum2009}. It was shown that as we increase the frequency of the oscillation, the overlap decreases between the template and these non-BD bispectra. Also the proposed template does not have the correct scaling in the squeezed limit, making it less useful for for example bias in large scale structure.  

\begin{figure}
\resizebox{0.92\hsize}{!}{
\includegraphics{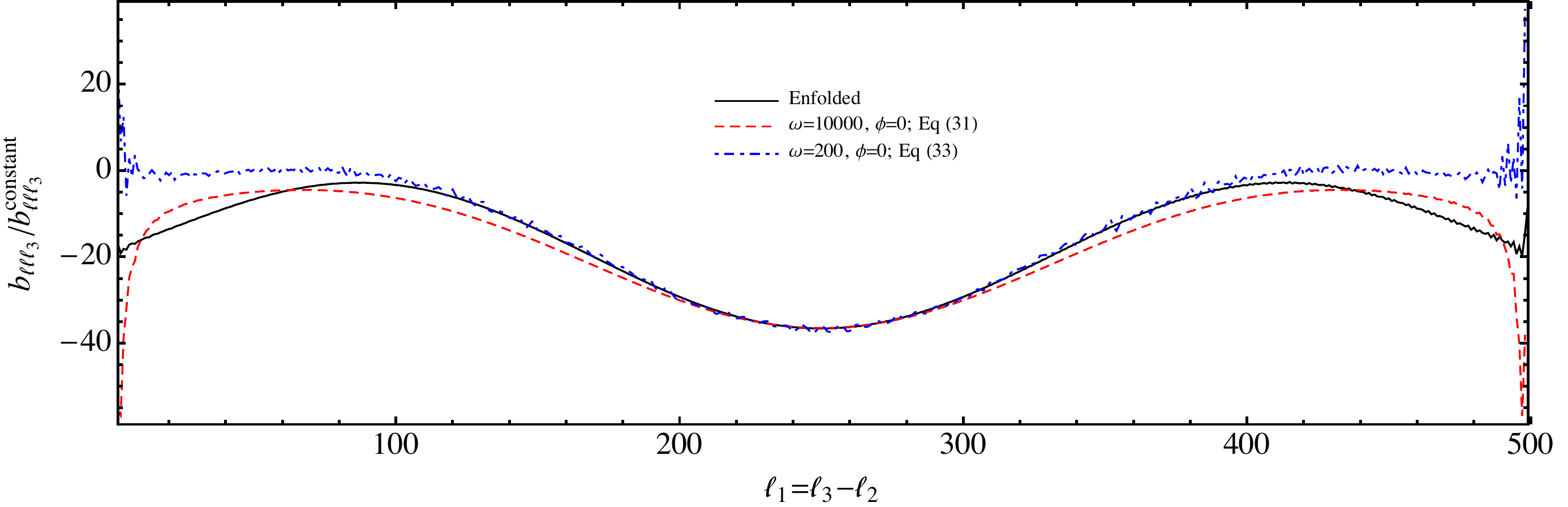}
}
\caption{The enfolded template as proposed in \cite{NonBDBispectrum2009} compared to the shape of Eq.~\eqref{eq:simplebd} and Eq.~\eqref{eq:lesssimplebd}. It has strong visual resemblance and a more quantitative comparison reveals that these shapes overlap $97-99\%$ in this limit. It also shows that although the enfolded captures the collinear limit, it does worse in the squeezed limit, where the visual resemblance is less apparent. However, since it is finite in this limit, it does not ruin the total overlap.  }
\label{fig:Enfolded_vs_nonBD_shape1}
\end{figure}

The CMB multipole correlator between two bispectra $B$ and $B'$ is given by
\be
{\rm Corr}(B,B') = \frac{1}{{\rm norm}(B){\rm norm}(B')} \sum_{\ell} \frac{B_{\ell_1\ell_2 \ell_3}B'_{\ell_1 \ell_2 \ell_3}}{\mathcal{C}_{\ell_1}\mathcal{C}_{\ell_3}\mathcal{C}_{\ell_3}}
\ee 
We first consider the collinear limit, before calculating the full (all $\ell$) correlator below. To show that the enfolded template is in fact useful in the collinear limit, we have computed the overlap in the collinear limit between the enfolded template and Eq.~\eqref{eq:simplebd} on the collinear axis above (i.e. $\ell_3 = 500 = \ell_1+\ell_2$). We find that the overlap on the collinear axis increases from $80 \%$ for $\omega = 50$ to $99\%$ for $\omega=10000$. The reason why the overlap goes up is that as we increase the frequency the collinear enhancement becomes much more pronounced. In Fig.~\ref{fig:Enfolded_vs_nonBD_shape1} we show the collinear limit of the enfolded and non-BD shape of Eq.~\eqref{eq:simplebd} (with some arbitrary normalization). It is clear that the features in this limit are very similar, specifically the enhancement in the limit $\ell_1 = \ell_2$. It is also clear that this template does not do well in the squeezed limit (i.e. when $\ell_1 \ll \ell_2,\ell_3$), which is relevant for using the template for the computations of the non-Gaussian bias in large scale structure. We find similar overlap for the shape in Eq.~\eqref{eq:lesssimplebd}, with overlap almost independent of frequency $\sim 95-97\%$. 

\begin{figure}
\resizebox{0.92\hsize}{!}{
\includegraphics{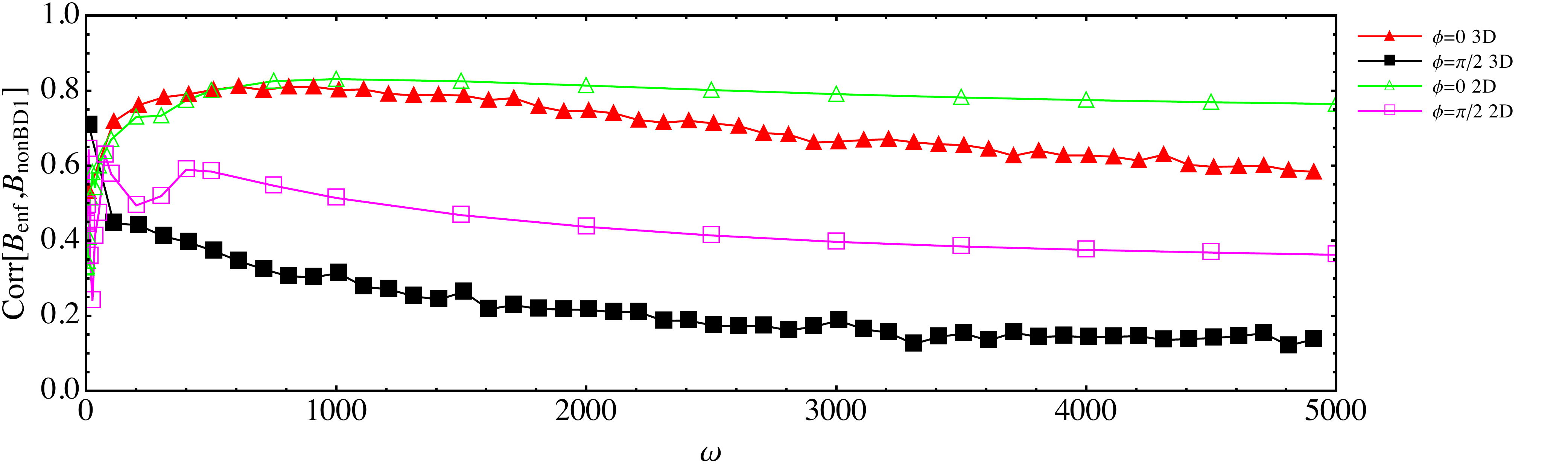}
}
\caption{Primordial (3D) and CMB multipole (2D) bispectrum correlation of the enfolded template and the exact non-BD shape Eq.~\eqref{eq:simplebd} as a function of frequency $\omega$ for 2 values of the phase. Although the overlap in the collinear limit matches well with the enfolded template, rapid oscillations off the collinear limit result in a poor overlap. After projection some of the oscillations wash out and the overlap improves. We also investigated the overlap with the second shape (Eq.~\eqref{eq:lesssimplebd}), and found much worse correlation due to the scale dependent phase, establishing a strong case for a more accurate reconstruction. }
\label{fig:overlap_enfolded}
\end{figure}

To show that there is a need for an improvement over just a simple enfolded template, we computed the overlap of the full shape up to $\ell_{\rm max}=2000$ with the enfolded template for various frequencies, both in  primordial $k$ space and in CMB multipole space. We use the primordial shape correlator \cite{NonGaussianShapes,NonBDBispectrum2009}
\be
F (S_1,S_2) \equiv \int d\mathcal{V} S_1(k_1,k_2,k_3)S_2(k_1,k_2,k_3) \mathcal{W}(k_1,k_2,k_3).
\ee
and ${\rm Corr}^{3D}(B,B')  = F(S_1,S_2)/(F(S_1)F(S_2))$ and the weight proposed in \cite{ShellardModeExpansion2009} $\mathcal{W}(k_1,k_2,k_3) = k_t^{-1}$. We show both the priordial and the CMB correlator between the enfolded template and Eq.~\eqref{eq:simplebd} as a function of $\omega$ and for two values of the phase in Fig.~\ref{fig:overlap_enfolded}. First of all, we find that as we increase the frequency the overlap decreases as was shown earlier Ref.~\cite{NonBDBispectrum2009}. Secondly, the enfolded shape is most adapted to a phase $\phi = 0$. Any deviation from that generally leads to a worse overlap, as was pointed out in Ref.~\cite{NonBDBispectrum2010}; since the phase is a free parameter, we would severely limit the extend of parameter space we are sensitive to if we would rely on the enfolded template only. Thirdly, after projection the overlap improves as oscillations are washed out. This was also found in Ref.~\cite{NonBDBispectrum2009}. 

It is worth pointing out that the primordial correlator is a function of $k_{\rm max}$. Even though our $k_{\rm max}$ is set to 0.7 Mpc$^{-1}$ for the CMB multipole bispectra, in order to obtain somewhat reliable overlap in primordial space, the primordial correlator has to be cutoff around the Silk damping scale $k_{\rm max}\sim k_{\rm silk}=0.1$ Mpc$^{-1}$. Depending on their functional form, some oscillating shapes are less sensitive to the value of $k_{\rm max}$. One should be aware of this difference when comparing the results presented here with those found in Ref.~\cite{NonBDBispectrum2009}; there the shape considered has a scale dependent cutoff. Consequently, the overlap did not depend on $k_{\rm max}$ and the effective frequencies are somewhat higher and therefore the overlap smaller. In addition, for the more complicated shape of Eq.~\eqref{eq:lesssimplebd}, there is a $k$-dependent phase. As a result of the scale dependent phase of this shape we find a very small overlap ($<$ a few percent) for most frequencies and phases. We only investigated the overlap of this shape in primordial space, but although some of the oscillations will be washed out by projection, we do not expect large improvements of the correlator in CMB multipole space. 

Our aim is to squeeze every bit of information out of the CMB and therefore we prefer a representation of the theoretical shape that is $>99\%$. For that purposes we require a modal expansion, which will be the topic of the next section. 

\subsection{Mode decomposition}

As we we have seen above, non BD bispectrum shapes are generally of the form $B(k_t,K_j)$, with appropriate prefactors of single momenta $k_i$ that are not an obstacle for separability. Our goal is to find efficient expansions of our primordial shapes in terms of separable basis functions, exploiting the symmetries of the primordial shapes. This will allow us to construct efficient estimators and make signal-to-noise forecasts in the following sections.

\subsubsection{Effectively 1-dimensional shapes}

The simplest shape of interest is Eq.~\eqref{eq:simplebd} which is only a function of $K_j$. In this case, we can use a 1-dimensional expansion similar to that in Ref.~\cite{OptimalEstimator2014}. In general any function $f(x)$ can be expanded in a Fourier series $\sum_m \alpha_m e^{imx}$, where $x$ in the present case is some function of the momenta $k_i$. The resulting expansion will be useful for our purpose if the exponentials are separable, i.e. if $x$ is a linear function of momenta $k_i$ and not a more complicated function of these. Therefore we can expand Eq.~\eqref{eq:simplebd} as 
\be
B(K_1,K_2,K_3) = \sum_{j=1}^3 \sum_{n=-N}^{N} c_n e^{i\frac{2\pi n K_j}{\Delta K_j}}, 
\ee
with expansion coefficients
\be
c_n = \frac{1}{\Delta K_j} \int_{K_j^{min}}^{K_j^{max}} B(K_j) e^{-i\frac{2\pi n K_j}{\Delta K_j}} dK_j.
\ee

\begin{figure}
\resizebox{0.7\hsize}{!}{
\includegraphics{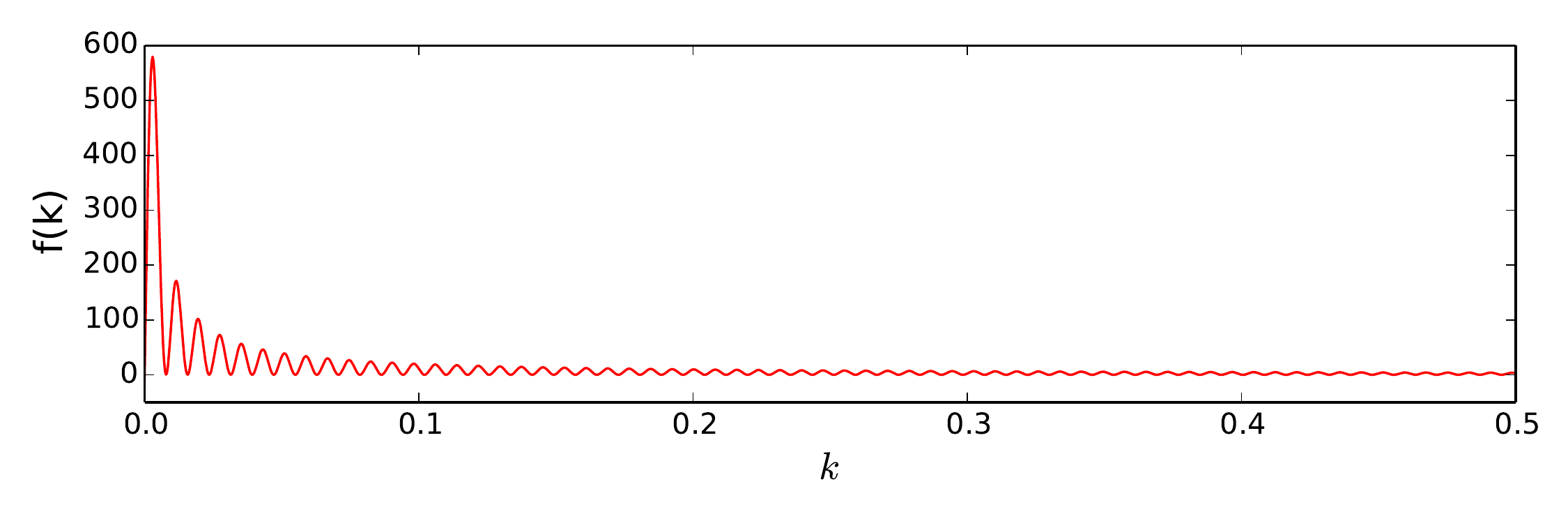}
}
\resizebox{0.7\hsize}{!}{
\includegraphics{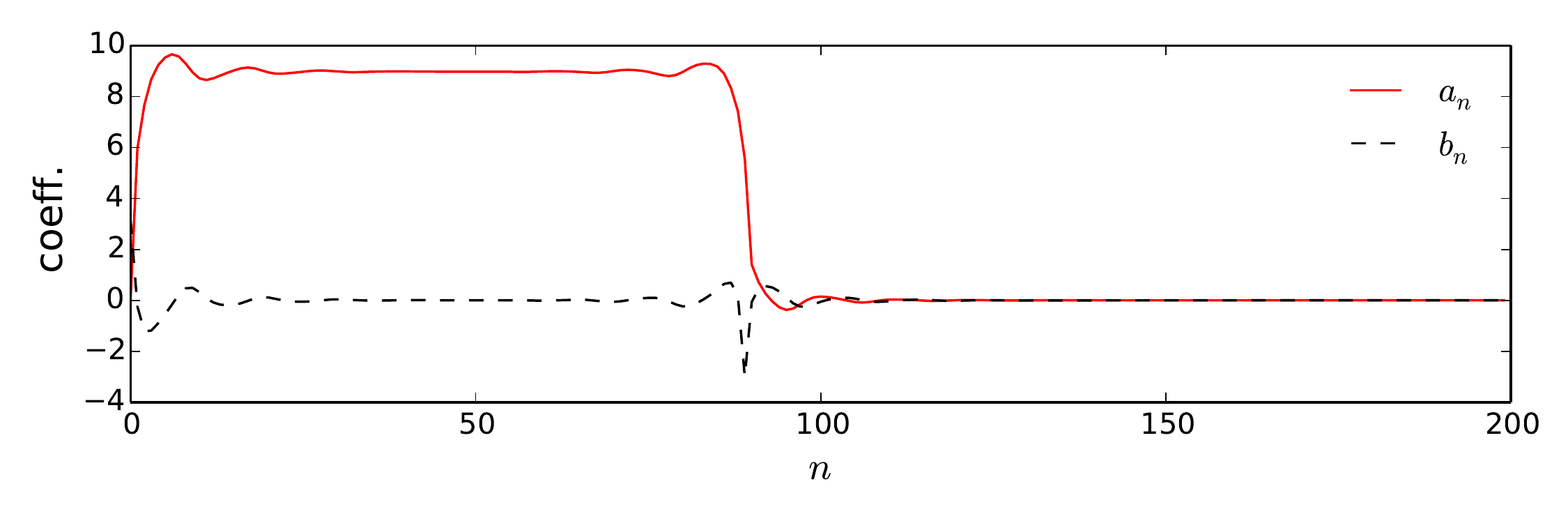}
}
\caption{Primordial shape $f(K_j) = \frac{\cos(\omega K_j)}{K_j}$ (top) and corresponding modal coefficients $c_n = a_n + i b_n$ (bottom) for $\omega=800$. }
\label{fig:modalcoeff_1dkj}
\end{figure}

The range of integration over the 1-dimensional variable $K_j$ corresponds to the physically allowed values of $K_j$, i.e. $0<K_j<2 k_{max}$, where $k_{max}$ is the maximal single momentum for which we want to describe the primordial bispectrum. The bispectrum must be periodic on the space of integration, which we achieve by multiplying with a generalized gaussian window function (see Ref.~\cite{OptimalEstimator2014}). Fig.~\ref{fig:modalcoeff_1dkj} shows an example expansion of the primordial bispectrum in Eq.~\eqref{eq:simplebd} with $(\omega=800, \phi=0)$. We see that our expansion in terms of oscillating basis functions is well suited to represent both the fast oscillations and the enhancement for $K_j \rightarrow 0$.

\begin{figure}
\resizebox{0.9\hsize}{!}{
\includegraphics{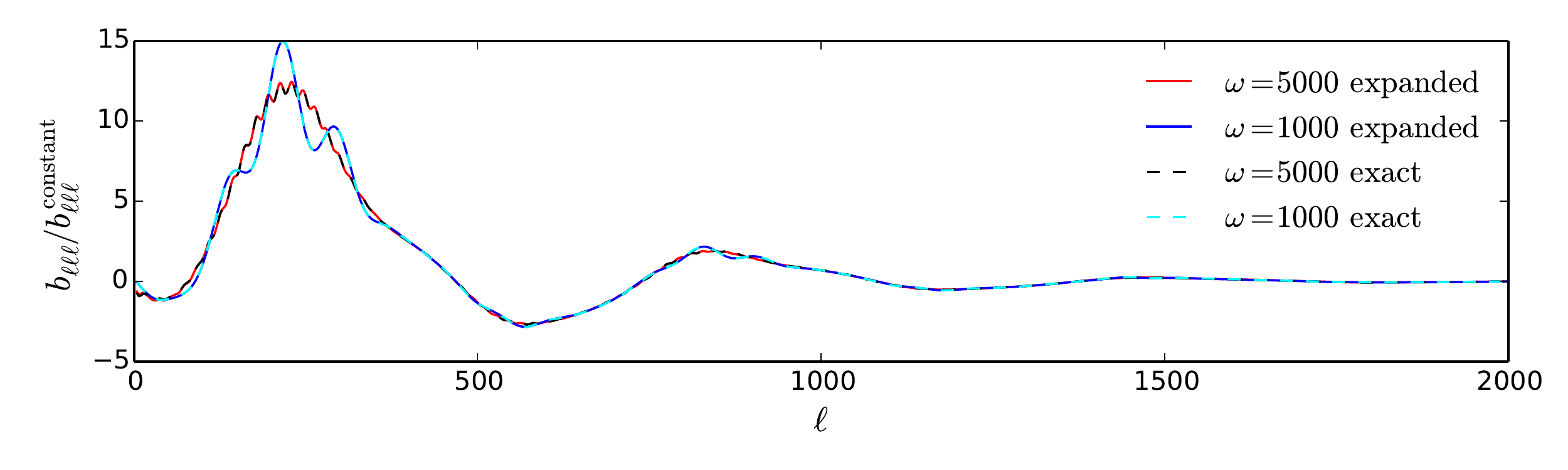}
}
\resizebox{0.9\hsize}{!}{
\includegraphics{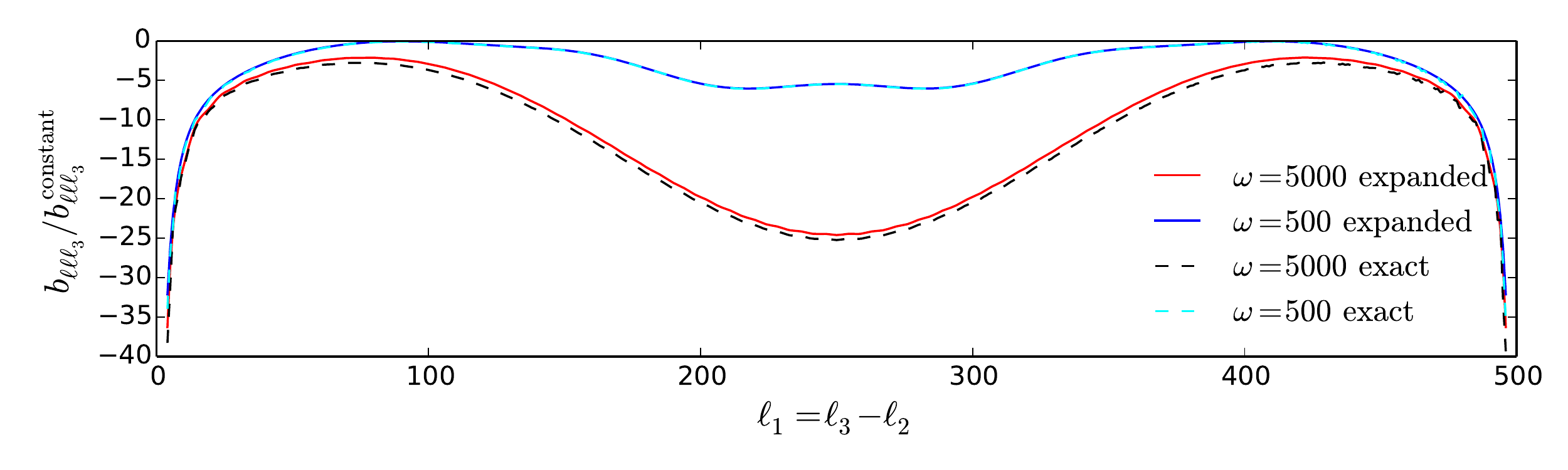}
}
\caption{Comparison of the exact bispectrum results from section \ref{sec:exactcalc} and the expansion proposed here, using 600 modes. Top: Equilateral limit $\ell_1 = \ell_2 = \ell_3$. Bottom: Collinear limit $\ell_3 = 500 = \ell_1+\ell_2$. The results coincide as expected.}
\label{fig:bispectrum_compare_exact_expansion}
\end{figure}

The CMB multipole bispectrum for the shape in Eq.~\eqref{eq:simplebd} is given by
\be
\label{eq:blll_nonbd1}
b_{l_1l_2l_3} = \int dx x^2 \sum_{n=-N}^N c_{n} \left( A^{l1}_{n}(x)A^{l2}_{n}(x)B^{l3}_{-n}(x) + A^{l1}_{n}(x)B^{l2}_{-n}(x)A^{l3}_{n}(x) +B^{l1}_{-n}(x)A^{l2}_{n}(x)A^{l3}_{n}(x) \right),
\ee
where
\be
\label{eq:kswfilter1}
A^{\ell}_n(x) &= \frac{2}{\pi} \int dk k^2 A_n(k) j_{\ell}(kx) \Delta_{\ell}(k), \hspace{1cm}
B^{\ell}_n(x) &= \frac{2}{\pi} \int dk k^2 B_n(k) j_{\ell}(kx) \Delta_{\ell}(k),
\ee
and $A_n(k) = \frac{e^{ink}}{k}$ and $B_n(k) = \frac{e^{ink}}{k^3}$. Fig.~\ref{fig:bispectrum_compare_exact_expansion} shows the bispectrum calculated with our expansion using 600 modes compared to the exact result of section \ref{sec:exactcalc}.

\subsubsection{Effectively 2-dimensional shapes}
For shapes $B(k_t,K_j)$ like Eq.~\eqref{eq:shape_nonbd2} we construct an efficient 2-dimensional modal expansion. The general 2-dimensional Fourier expansion, of a function $f(x,y)$ where $x$ and $y$ are two independent variables, is 
\be
f(x,y) = \sum_{mn} c_{mn} e^{i \frac{2 \pi m x}{\Delta x}}e^{i \frac{2 \pi n y}{\Delta y}}.
\ee
We can therefore expand the bispectrum as
\be
B(k_t,K_1,K_2,K_3) = \sum_j \sum_{mn} c_{mn} e^{i\frac{2 \pi m K_j}{\Delta K_j}} e^{i \frac{2 \pi n k_t}{\Delta k_t}}.
\ee
The Fourier coefficients are given by the 2-dimensional integral 
\be
c_{mn}  = \frac{1}{\Delta K_j \Delta k_t} \int_{K_j^{min}}^{K_j^{max}} \int_{k_t^{min}}^{k_t^{max}} B(K_j,k_t) e^{-i\frac{2\pi n K_j}{\Delta K_j}} e^{-i\frac{2\pi n k_t}{\Delta k_t}}  dK_j dk_t,
\ee
where $0<K_j<2 k_{max}$ and $3 k_{min}<k_t<3 k_{max}$. Defining $m' = \frac{m}{\Delta K_j}$ and $n' = \frac{n}{\Delta k_t}$ and expressing the bispectrum as a function of single momenta $k_i$, thereby exploiting the relation of $k_t$ and $K_j$, we find
\be
B(k_1,k_2,k_3) = \sum_{mn} c_{mn} \big[A_{m'+n'}(k_1)A_{m'+n'}(k_2)B_{-m'+n'}(k_3) &+& A_{m'+n'}(k_1)B_{-m'+n'}(k_2)A_{m'+n'}(k_3) \\ \nonumber
&+& B_{-m'+n'}(k_1)A_{m'+n}(k_2)A_{m'+n'}(k_3) \big],
\ee
where 
$A_a(k) = \frac{e^{i 2 \pi a k}}{k^s}$ and $B_a(k) = \frac{e^{i 2 \pi a k}}{k^t }$. Here $s,t$ are the $k$ scalings that have to be adjusted to the exact shape under consideration. 

The CMB multipole bispectrum is then given by
\be
b_{l_1l_2l_3} = \int dx x^2 \sum_{mn} c_{mn}  A^{l1}_{m'+n'}(x)A^{l2}_{m'+n'}(x)B^{l3}_{-m'+n'}(x) + \left( \mathrm{perm. \,+m \leftrightarrow -m} \right).
\ee
The time critical part of the CMB bispectrum is the calculation of the functions $A^{l}_{n}(x)$ and $B^{l}_{n}(x)$. By choosing the size of the integration domains $\Delta K_j$ and $\Delta k_t$ identical or a multiple of each other, one can reduce the number of functions that have to be calculated considerably. While our 2-dimensional expansion is computationally more demanding than a 1-dimensional one, it is still a considerable improvement with respect to a general 3-dimensional modal expansion for arbitrary shapes. In addition, the use of oscillating basis functions makes it optimally suitable for the oscillating shapes created by non-Bunch Davies initial conditions.

\subsection{KSW type modal estimator}\label{sec:estimator}

The KSW estimator for the 1-dimensional expansion in $K_j$ is similar to the one for $k_t$ in Ref.~\cite{OptimalEstimator2014}. It is given by $\hat{f} = \frac{1}{N} \left(S_{\rm cub}+S_{\rm lin}\right)$ where, using complex quantities, the cubic term is 
\be
S^{\rm cub} = \sum_{n} c_{n} \int r^2 dr \int d\Omega  A^2_{n}(r,\hat{n})B_{-n}(r,\hat{n}), 
\ee
and the linear correction (to take into account noise and partial sky coverage) is
\be
S^{\rm lin} = -3 \sum_{n} c_{n} \int r^2 dr \int d\Omega  \bigl[ B_n(r,\hat{n}) \left< A^2_n(r,\hat{n}) \right> + 2 A_n(r,\hat{n}) \left< B_n(r,\hat{n}) A_n(r,\hat{n}) \right> \bigr],
\ee
where $\left<... \right>$ are the usual Monte Carlo averages over simulated gaussian maps.
For the shape in equation Eq.~\eqref{eq:simplebd} the filtered maps are given by 
\be
A_n(r, \hat{n}) =& \sum_{\ell m} (C^{-1}a)_{\ell m}A_n^{\ell}(r) Y_{\ell m}(\hat{n}), \hspace{1cm}
B_n(r, \hat{n}) =& \sum_{\ell m} (C^{-1}a)_{\ell m}B_n^{\ell}(r) Y_{\ell m}(\hat{n}),
\ee
with $A_n^{\ell}(r)$, $B_n^{\ell}(r)$ given in Eq.~\eqref{eq:kswfilter1}. The norm $N$ of the estimator will be calculated in the next section. The expansion factors $c_{n}$ must be evaluated for a sufficiently dense sampling in $\omega$ and $\phi$.

For the 2-dimensional expansion, the estimator is given by
\be
S^{\rm cub} = \sum_{mn} c_{mn} \int r^2 dr \int d\Omega  A^2_{m'+n'}(r,\hat{n})B_{-m'+n'}(r,\hat{n}),  
\ee
and 
\be
S^{\rm lin} = -3 \sum_{mn} c_{mn} \int r^2 dr \int d\Omega  \bigl[ B_{-m'+n'} (r,\hat{n}) \left< A^2_{m'+n'}(r,\hat{n}) \right> + 2 A_{m'+n'}(r,\hat{n}) \left< B_{-m'+n'} (r,\hat{n}) A_{m'+n'}(r,\hat{n}) \right> \bigr].
\ee
The computationally most demanding part of the KSW estimator, the computation of the filtered maps, does not change fundamentally when going from the $1d$ to the $2d$ expansion. Only the multiplication of these real space maps in the estimator scales quadratically with the number of modes. However this part is subdominant in the calculation, making the estimator computationally feasible.

\subsection{Fisher forecast}

\begin{figure}
\resizebox{0.9\hsize}{!}{
\includegraphics{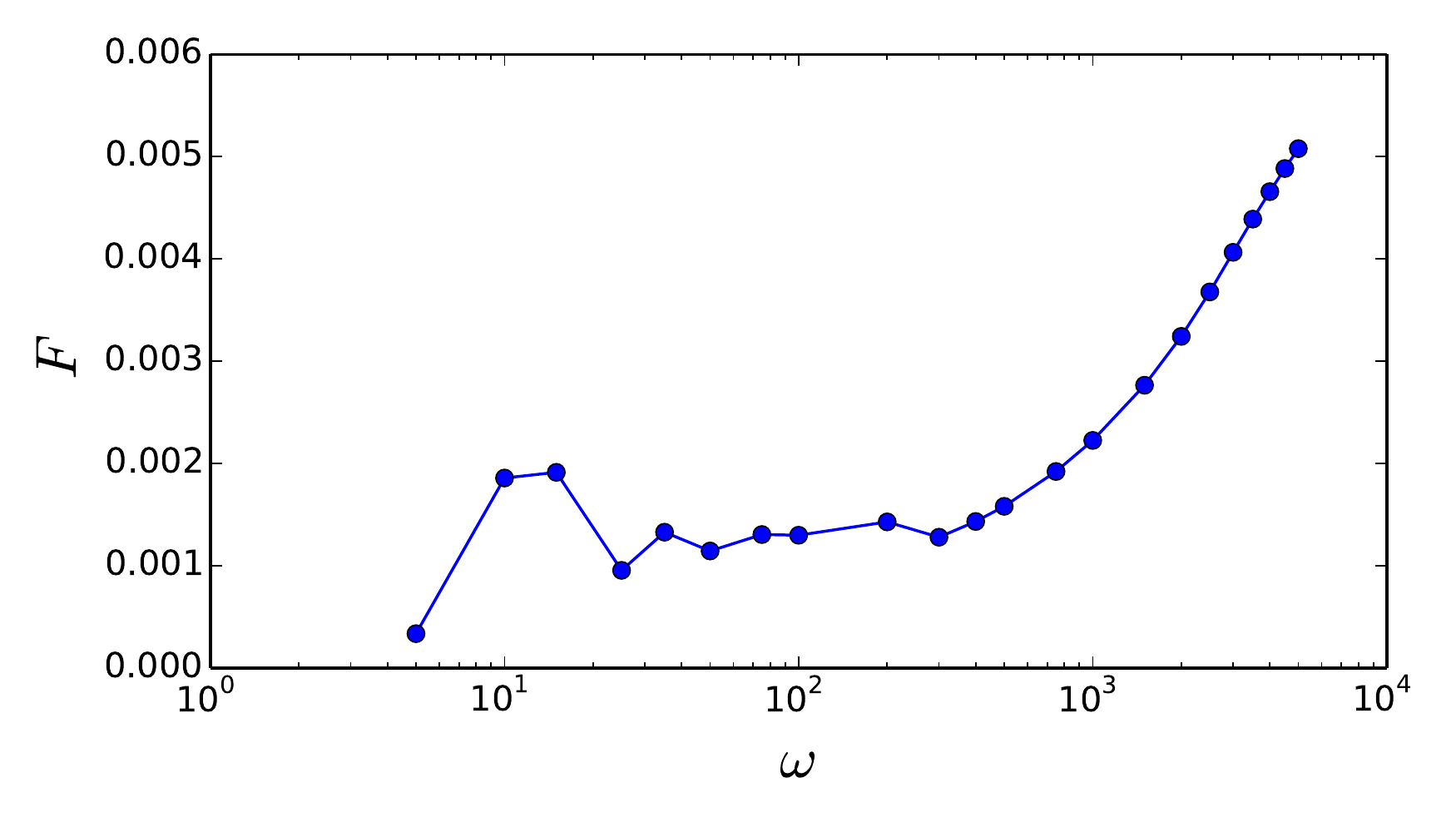}\includegraphics{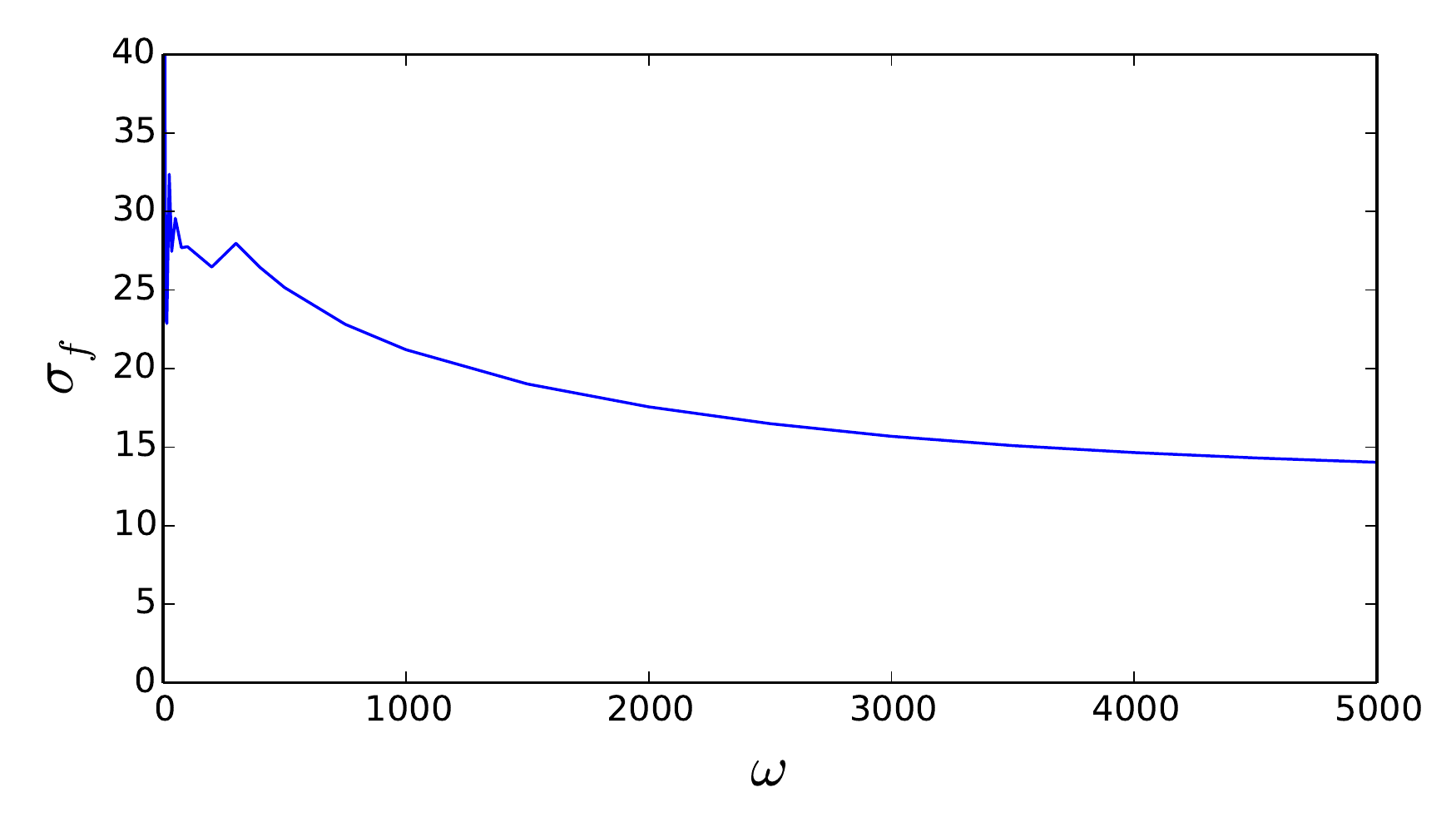}
}
\resizebox{0.5\hsize}{!}{
\includegraphics{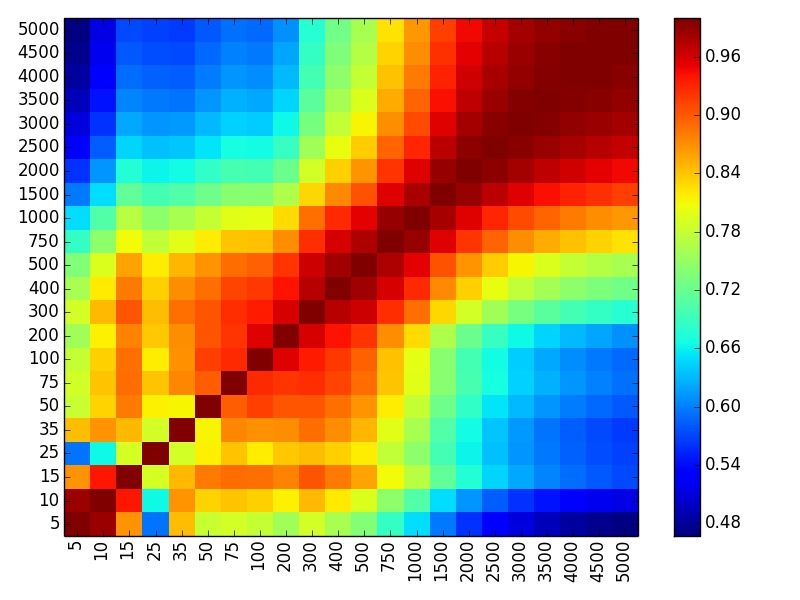}
}
\caption{Top: Fisher matrix value $F_{ii}$ (left) and corresponding standard deviation (right) as a function of frequency $\omega$. Bottom: Correlation matrix ${\rm corr}(f_i,f_j)$ as a function of frequencies $\omega_i,\omega_j$.}
\label{fig:fisher1}
\end{figure}

In this section we calculate the Fisher matrix, which provides the signal-to-noise ratio as well as the normalization of the KSW estimator, for the simple 1-dimensional shape in Eq.~\eqref{eq:simplebd}. For a CMB experiment with noise power spectrum $N_l$ and sky coverage fraction $f_{sky}$, the Fisher matrix for bispectra indexed by $i,j$ (here discriminating between frequencies $\omega_i,\omega_j$) is 
\be
\label{eq_fisher3} 
F_{ij} = f_{sky} \sum_{l_1l_2l_3}  I_{\ell_1\ell_2\ell_3}  \frac{ b^i_{\ell_1\ell_2\ell_3} b^j_{\ell_1\ell_2\ell_3} }{C_{\ell_1}C_{\ell_2}C_{\ell_3}},
\ee
where the reduced CMB multipole bispectrum $b_{l_1l_2l_3}$ is here given by eq. \ref{eq:blll_nonbd1}, $C_l=C_l^{CMB}+N_l$ is the power spectrum, and $I_{\ell_1\ell_2\ell_3}$ is the usual geometric factor. We normalize the bispectrum in analogy with other oscillating shapes in the literature as
\begin{equation}
\label{eq_oscispectrum1} 
B_\Phi(k_1,k_2,k_3) = \frac{ \Delta_\Phi^2 f_{\rm NL}}{(k_1k_2k_3)^2} \sum_j\frac{1-\cos \omega K_j}{k_j^2K_j},
\end{equation}
and we assume $\Delta_\Phi^2 = 9.04\times10^{-16}$. The diagonal Fisher matrix value $F_{ii}$ as well as the corresponding standard deviation $\sigma_f = \frac{1}{\sqrt{F_{ii}}}$ of an optimal measurement with the estimator of the previous section is plotted in figure \ref{fig:fisher1} (top). For low $\omega$, the cosine is nearly constant and therefore the bispectrum Eq.~\eqref{eq_oscispectrum1} goes to zero. From the Fisher matrix one can also calculate the correlation coefficient ${\rm corr}(f_i,f_j) = \frac{F_{ij}}{\sqrt{F_{ii} F_{jj}} }$, which is plotted in Fig.~\ref{fig:fisher1} (bottom) for some test frequencies. It gives an impression of the necessary sampling in frequency space when doing the estimation. As one might expect from the visualization of spectra in previous sections, there is a large overlap in wide ranges of frequencies.

For the resonant bispectrum \cite{OptimalEstimator2014} $\sigma$ increased as a function of $\omega$. For the resonant bispectrum, the increasing error is caused by the rapid oscillations. For the bispectrum considered we also expect rapid oscillations (off-collinear), but the produced bispectra from excited states are naturally enhanced in the collinear limit with and enhancement linear in $\omega$ for the bispectrum of Eq.~\eqref{eq:simplebd}. If the scaling would hold after projection, we would expect that $\sigma$ would drop accordingly, so as $1/\omega$. However, as was pointed out in \cite{NonBDBispectrum2009}, the presence of oscillations and because of projection one looses some of this enhancement. It was argued through an analytic computation that one would in fact loose the full enhancement $\omega$ in Ref. \cite{InitialStateOriginalHolman2007}, but here and in Ref.~\cite{NonBDBispectrum2009}, we roughly find that $\sigma \propto 1/ \log \omega$. In other words, the error decreases as you increase the number of oscillations. 

For realistic values of $(H/M_{\rm pl})$ this shape is most likely unobservable in the CMB. An exception would only occur if the fundamental scale of the underlying high energy theory is similar to the energy scale of inflation, and well below the Planck scale, which increases $\omega$ and thus the remaining enhancement. Our main objective in this paper was to find suitable expansions for the typical $k$-dependencies of non-BD shapes, and the present shape is merely an illustration of the most simple case. More complicated Hamiltonians with higher derivative operators would lead to potentially observable signals \cite{InitialStateOriginalHolman2007,NonBDBispectrum2009,NonBDBispectrum2010,EFTexcitedStates}, even for small values of $\omega$. Although the scale dependence of these bispectra is more complicated, the overall scaling is preserved, including the presence of features and enhancement in the collinear limit. The tools presented here to reconstruct the simplest case, are straightforwardly extended for more complicated shapes and we will include the search for these spectra in the data in future work.

\section{Conclusions}\label{sec:conclusion}

In this paper we developed an optimal set of estimators for bispectra from excited states during inflation. Although the detailed scale dependence of the bispectra can differ based on several assumptions, one can distinguish two key properties that are very general; first of all, the predicted bispectra are enhanced in the collinear limit, and  secondly, the bispectra contain oscillating features. Both of these phenomenological properties make it hard to use existing techniques to build fast, optimal estimators. Similar to the reconstruction of the axion bispectrum in \cite{OptimalEstimator2014}, we used the fact that all these bispectra are effectively at most 2-dimensional, allowing for a much faster reconstruction using a simple 1 or 2-dimensional Fourier basis. We tested our expansion, and showed that we can achieve good overlap with the exact result, using a limited number of modes, even for highly oscillating bispectra. 

We also calculated the overlap with the enfolded template, which has been used to search for bispectra from excited states in the CMB. We found that in the collinear limit the predicted bispectra have large overlap with the enfolded template. In Ref.~\cite{NonBDBispectrum2009} it was shown that the overlap decreases as a function of the frequency of the bispectrum (for an NPH pure state rotation); we find that although this is true, in the collinear limit itself, the overlap generally increases as a function of frequency. As was pointed out in e.g. Ref. \cite{BiasVerde2010}, the enfolded template does not have the correct scaling in the local limit (making it less useful when considering effects on the halo bias), however the squeezed contribution does not diverge. The enfolded template has poor overlap anywhere else, in particular when the bispectrum has a non-vanishing phase. This establishes a strong case for the reconstruction approach presented in this paper. 

The ultimate experiment combines measurement of all the $n$-point correlation functions. Hence, in this paper we have stressed the need for consistency when predicting effects from excited states. Although the differences are typically of order slow-roll in the power spectrum, at the very least we should be aware of them. For the bispectrum they can lead to much larger differences and it is crucial that the correct estimator is applied when searching for the associated signal of a given model. 

Together with the optimal estimator for the resonant bispectrum, we have now completed a framework that allows us to constrain a large class of bispectra that were difficult to constrain with existing methods. In this paper we focused on a very simple model, with only a single field and no non-canonical kinetic terms and a quadratic potential. In a more complete picture \cite{NonBDBispectrum2010,NonBDBispectrum2010b}, more complicated bispectra are predicted. The framework presented here can easily be extended to capture those shapes as well. We hope to report on this search in future work. 

\section*{Acknowledgments} We would like to thank Francois Bouchet, Robert Brandenberger, Guido D'Amico, Dan Green, Matt Kleban, Jerome Martin, Jan Pieter van der Schaar, David Spergel and Benjamin Wandelt for useful discussions and comments. MM acknowledges funding by Centre National d'Etudes Spatiales (CNES).

\bibliography{MDB}

\begin{thebibliography}{50}
\expandafter\ifx\csname natexlab\endcsname\relax\def\natexlab#1{#1}\fi
\expandafter\ifx\csname bibnamefont\endcsname\relax
  \def\bibnamefont#1{#1}\fi
\expandafter\ifx\csname bibfnamefont\endcsname\relax
  \def\bibfnamefont#1{#1}\fi
\expandafter\ifx\csname citenamefont\endcsname\relax
  \def\citenamefont#1{#1}\fi
\expandafter\ifx\csname url\endcsname\relax
  \def\url#1{\texttt{#1}}\fi
\expandafter\ifx\csname urlprefix\endcsname\relax\def\urlprefix{URL }\fi
\providecommand{\bibinfo}[2]{#2}
\providecommand{\eprint}[2][]{\url{#2}}

\bibitem[{\citenamefont{{Planck Collaboration}
  et~al.}(2015{\natexlab{a}})\citenamefont{{Planck Collaboration}, {Ade},
  {Aghanim}, {Arnaud}, {Ashdown}, {Aumont}, {Baccigalupi}, {Banday},
  {Barreiro}, {Bartlett} et~al.}}]{PlanckCosmoPars2015}
\bibinfo{author}{\bibnamefont{{Planck Collaboration}}},
  \bibinfo{author}{\bibfnamefont{P.~A.~R.} \bibnamefont{{Ade}}},
  \bibinfo{author}{\bibfnamefont{N.}~\bibnamefont{{Aghanim}}},
  \bibinfo{author}{\bibfnamefont{M.}~\bibnamefont{{Arnaud}}},
  \bibinfo{author}{\bibfnamefont{M.}~\bibnamefont{{Ashdown}}},
  \bibinfo{author}{\bibfnamefont{J.}~\bibnamefont{{Aumont}}},
  \bibinfo{author}{\bibfnamefont{C.}~\bibnamefont{{Baccigalupi}}},
  \bibinfo{author}{\bibfnamefont{A.~J.} \bibnamefont{{Banday}}},
  \bibinfo{author}{\bibfnamefont{R.~B.} \bibnamefont{{Barreiro}}},
  \bibinfo{author}{\bibfnamefont{J.~G.} \bibnamefont{{Bartlett}}},
  \bibnamefont{et~al.}, \bibinfo{journal}{ArXiv e-prints}
  (\bibinfo{year}{2015}{\natexlab{a}}), \eprint{1502.01589}.

\bibitem[{\citenamefont{{Bennett} et~al.}(2012)\citenamefont{{Bennett},
  {Larson}, {Weiland}, {Jarosik}, {Hinshaw}, {Odegard}, {Smith}, {Hill},
  {Gold}, {Halpern} et~al.}}]{WMAPfinal2012}
\bibinfo{author}{\bibfnamefont{C.~L.} \bibnamefont{{Bennett}}},
  \bibinfo{author}{\bibfnamefont{D.}~\bibnamefont{{Larson}}},
  \bibinfo{author}{\bibfnamefont{J.~L.} \bibnamefont{{Weiland}}},
  \bibinfo{author}{\bibfnamefont{N.}~\bibnamefont{{Jarosik}}},
  \bibinfo{author}{\bibfnamefont{G.}~\bibnamefont{{Hinshaw}}},
  \bibinfo{author}{\bibfnamefont{N.}~\bibnamefont{{Odegard}}},
  \bibinfo{author}{\bibfnamefont{K.~M.} \bibnamefont{{Smith}}},
  \bibinfo{author}{\bibfnamefont{R.~S.} \bibnamefont{{Hill}}},
  \bibinfo{author}{\bibfnamefont{B.}~\bibnamefont{{Gold}}},
  \bibinfo{author}{\bibfnamefont{M.}~\bibnamefont{{Halpern}}},
  \bibnamefont{et~al.}, \bibinfo{journal}{ArXiv e-prints}
  (\bibinfo{year}{2012}), \eprint{1212.5225}.

\bibitem[{\citenamefont{{Das} et~al.}(2014)\citenamefont{{Das}, {Louis},
  {Nolta}, {Addison}, {Battistelli}, {Bond}, {Calabrese}, {Crichton}, {Devlin},
  {Dicker} et~al.}}]{ACT2014}
\bibinfo{author}{\bibfnamefont{S.}~\bibnamefont{{Das}}},
  \bibinfo{author}{\bibfnamefont{T.}~\bibnamefont{{Louis}}},
  \bibinfo{author}{\bibfnamefont{M.~R.} \bibnamefont{{Nolta}}},
  \bibinfo{author}{\bibfnamefont{G.~E.} \bibnamefont{{Addison}}},
  \bibinfo{author}{\bibfnamefont{E.~S.} \bibnamefont{{Battistelli}}},
  \bibinfo{author}{\bibfnamefont{J.~R.} \bibnamefont{{Bond}}},
  \bibinfo{author}{\bibfnamefont{E.}~\bibnamefont{{Calabrese}}},
  \bibinfo{author}{\bibfnamefont{D.}~\bibnamefont{{Crichton}}},
  \bibinfo{author}{\bibfnamefont{M.~J.} \bibnamefont{{Devlin}}},
  \bibinfo{author}{\bibfnamefont{S.}~\bibnamefont{{Dicker}}},
  \bibnamefont{et~al.}, \bibinfo{journal}{\jcap} \textbf{\bibinfo{volume}{4}},
  \bibinfo{eid}{014} (\bibinfo{year}{2014}), \eprint{1301.1037}.

\bibitem[{\citenamefont{{Keisler} et~al.}(2011)\citenamefont{{Keisler},
  {Reichardt}, {Aird}, {Benson}, {Bleem}, {Carlstrom}, {Chang}, {Cho},
  {Crawford}, {Crites} et~al.}}]{SPT2011}
\bibinfo{author}{\bibfnamefont{R.}~\bibnamefont{{Keisler}}},
  \bibinfo{author}{\bibfnamefont{C.~L.} \bibnamefont{{Reichardt}}},
  \bibinfo{author}{\bibfnamefont{K.~A.} \bibnamefont{{Aird}}},
  \bibinfo{author}{\bibfnamefont{B.~A.} \bibnamefont{{Benson}}},
  \bibinfo{author}{\bibfnamefont{L.~E.} \bibnamefont{{Bleem}}},
  \bibinfo{author}{\bibfnamefont{J.~E.} \bibnamefont{{Carlstrom}}},
  \bibinfo{author}{\bibfnamefont{C.~L.} \bibnamefont{{Chang}}},
  \bibinfo{author}{\bibfnamefont{H.~M.} \bibnamefont{{Cho}}},
  \bibinfo{author}{\bibfnamefont{T.~M.} \bibnamefont{{Crawford}}},
  \bibinfo{author}{\bibfnamefont{A.~T.} \bibnamefont{{Crites}}},
  \bibnamefont{et~al.}, \bibinfo{journal}{\apj} \textbf{\bibinfo{volume}{743}},
  \bibinfo{eid}{28} (\bibinfo{year}{2011}), \eprint{1105.3182}.

\bibitem[{\citenamefont{{Reichardt} et~al.}(2012)\citenamefont{{Reichardt},
  {Shaw}, {Zahn}, {Aird}, {Benson}, {Bleem}, {Carlstrom}, {Chang}, {Cho},
  {Crawford} et~al.}}]{SPT2012}
\bibinfo{author}{\bibfnamefont{C.~L.} \bibnamefont{{Reichardt}}},
  \bibinfo{author}{\bibfnamefont{L.}~\bibnamefont{{Shaw}}},
  \bibinfo{author}{\bibfnamefont{O.}~\bibnamefont{{Zahn}}},
  \bibinfo{author}{\bibfnamefont{K.~A.} \bibnamefont{{Aird}}},
  \bibinfo{author}{\bibfnamefont{B.~A.} \bibnamefont{{Benson}}},
  \bibinfo{author}{\bibfnamefont{L.~E.} \bibnamefont{{Bleem}}},
  \bibinfo{author}{\bibfnamefont{J.~E.} \bibnamefont{{Carlstrom}}},
  \bibinfo{author}{\bibfnamefont{C.~L.} \bibnamefont{{Chang}}},
  \bibinfo{author}{\bibfnamefont{H.~M.} \bibnamefont{{Cho}}},
  \bibinfo{author}{\bibfnamefont{T.~M.} \bibnamefont{{Crawford}}},
  \bibnamefont{et~al.}, \bibinfo{journal}{\apj} \textbf{\bibinfo{volume}{755}},
  \bibinfo{eid}{70} (\bibinfo{year}{2012}), \eprint{1111.0932}.

\bibitem[{\citenamefont{{Alvarez} et~al.}(2014)\citenamefont{{Alvarez},
  {Baldauf}, {Bond}, {Dalal}, {de Putter}, {Dor{\'e}}, {Green}, {Hirata},
  {Huang}, {Huterer} et~al.}}]{LSSnonGaussianity2014}
\bibinfo{author}{\bibfnamefont{M.}~\bibnamefont{{Alvarez}}},
  \bibinfo{author}{\bibfnamefont{T.}~\bibnamefont{{Baldauf}}},
  \bibinfo{author}{\bibfnamefont{J.~R.} \bibnamefont{{Bond}}},
  \bibinfo{author}{\bibfnamefont{N.}~\bibnamefont{{Dalal}}},
  \bibinfo{author}{\bibfnamefont{R.}~\bibnamefont{{de Putter}}},
  \bibinfo{author}{\bibfnamefont{O.}~\bibnamefont{{Dor{\'e}}}},
  \bibinfo{author}{\bibfnamefont{D.}~\bibnamefont{{Green}}},
  \bibinfo{author}{\bibfnamefont{C.}~\bibnamefont{{Hirata}}},
  \bibinfo{author}{\bibfnamefont{Z.}~\bibnamefont{{Huang}}},
  \bibinfo{author}{\bibfnamefont{D.}~\bibnamefont{{Huterer}}},
  \bibnamefont{et~al.}, \bibinfo{journal}{ArXiv e-prints}
  (\bibinfo{year}{2014}), \eprint{1412.4671}.

\bibitem[{\citenamefont{{M{\"u}nchmeyer}
  et~al.}(2015)\citenamefont{{M{\"u}nchmeyer}, {Meerburg}, and
  {Wandelt}}}]{OptimalEstimator2014}
\bibinfo{author}{\bibfnamefont{M.}~\bibnamefont{{M{\"u}nchmeyer}}},
  \bibinfo{author}{\bibfnamefont{P.~D.} \bibnamefont{{Meerburg}}},
  \bibnamefont{and} \bibinfo{author}{\bibfnamefont{B.~D.}
  \bibnamefont{{Wandelt}}}, \bibinfo{journal}{\prd}
  \textbf{\bibinfo{volume}{91}}, \bibinfo{eid}{043534} (\bibinfo{year}{2015}),
  \eprint{1412.3461}.

\bibitem[{\citenamefont{{Fergusson} and
  {Shellard}}(2007)}]{ShellardBispectrum2006}
\bibinfo{author}{\bibfnamefont{J.~R.} \bibnamefont{{Fergusson}}}
  \bibnamefont{and} \bibinfo{author}{\bibfnamefont{E.~P.~S.}
  \bibnamefont{{Shellard}}}, \bibinfo{journal}{\prd}
  \textbf{\bibinfo{volume}{76}}, \bibinfo{eid}{083523} (\bibinfo{year}{2007}),
  \eprint{astro-ph/0612713}.

\bibitem[{\citenamefont{{Fergusson} et~al.}(2010)\citenamefont{{Fergusson},
  {Liguori}, and {Shellard}}}]{ShellardModeExpansion2009}
\bibinfo{author}{\bibfnamefont{J.~R.} \bibnamefont{{Fergusson}}},
  \bibinfo{author}{\bibfnamefont{M.}~\bibnamefont{{Liguori}}},
  \bibnamefont{and} \bibinfo{author}{\bibfnamefont{E.~P.~S.}
  \bibnamefont{{Shellard}}}, \bibinfo{journal}{\prd}
  \textbf{\bibinfo{volume}{82}}, \bibinfo{eid}{023502} (\bibinfo{year}{2010}),
  \eprint{0912.5516}.

\bibitem[{\citenamefont{{Byun} and {Bean}}(2013)}]{Bean2013}
\bibinfo{author}{\bibfnamefont{J.}~\bibnamefont{{Byun}}} \bibnamefont{and}
  \bibinfo{author}{\bibfnamefont{R.}~\bibnamefont{{Bean}}},
  \bibinfo{journal}{\jcap} \textbf{\bibinfo{volume}{9}}, \bibinfo{eid}{026}
  (\bibinfo{year}{2013}), \eprint{1303.3050}.

\bibitem[{\citenamefont{{Byun} et~al.}(2015)\citenamefont{{Byun}, {Agarwal},
  {Bean}, and {Holman}}}]{JoyceShapes2015}
\bibinfo{author}{\bibfnamefont{J.}~\bibnamefont{{Byun}}},
  \bibinfo{author}{\bibfnamefont{N.}~\bibnamefont{{Agarwal}}},
  \bibinfo{author}{\bibfnamefont{R.}~\bibnamefont{{Bean}}}, \bibnamefont{and}
  \bibinfo{author}{\bibfnamefont{R.}~\bibnamefont{{Holman}}},
  \bibinfo{journal}{ArXiv e-prints}  (\bibinfo{year}{2015}),
  \eprint{1504.01394}.

\bibitem[{\citenamefont{{Meerburg}}(2010)}]{BispectrumOscillations2010}
\bibinfo{author}{\bibfnamefont{P.~D.} \bibnamefont{{Meerburg}}},
  \bibinfo{journal}{\prd} \textbf{\bibinfo{volume}{82}}, \bibinfo{eid}{063517}
  (\bibinfo{year}{2010}), \eprint{1006.2771}.

\bibitem[{\citenamefont{{Battefeld} and {Grieb}}(2011)}]{Battefeld2011}
\bibinfo{author}{\bibfnamefont{T.}~\bibnamefont{{Battefeld}}} \bibnamefont{and}
  \bibinfo{author}{\bibfnamefont{J.}~\bibnamefont{{Grieb}}},
  \bibinfo{journal}{\jcap} \textbf{\bibinfo{volume}{12}}, \bibinfo{eid}{003}
  (\bibinfo{year}{2011}), \eprint{1110.1369}.

\bibitem[{\citenamefont{{Behbahani} et~al.}(2012)\citenamefont{{Behbahani},
  {Dymarsky}, {Mirbabayi}, and {Senatore}}}]{EFTOscillations2011}
\bibinfo{author}{\bibfnamefont{S.~R.} \bibnamefont{{Behbahani}}},
  \bibinfo{author}{\bibfnamefont{A.}~\bibnamefont{{Dymarsky}}},
  \bibinfo{author}{\bibfnamefont{M.}~\bibnamefont{{Mirbabayi}}},
  \bibnamefont{and}
  \bibinfo{author}{\bibfnamefont{L.}~\bibnamefont{{Senatore}}},
  \bibinfo{journal}{\jcap} \textbf{\bibinfo{volume}{12}}, \bibinfo{eid}{036}
  (\bibinfo{year}{2012}), \eprint{1111.3373}.

\bibitem[{\citenamefont{{Planck Collaboration}
  et~al.}(2015{\natexlab{b}})\citenamefont{{Planck Collaboration}, {Ade},
  {Aghanim}, {Arnaud}, {Arroja}, {Ashdown}, {Aumont}, {Baccigalupi},
  {Ballardini}, {Banday} et~al.}}]{PlanckInflation2014}
\bibinfo{author}{\bibnamefont{{Planck Collaboration}}},
  \bibinfo{author}{\bibfnamefont{P.~A.~R.} \bibnamefont{{Ade}}},
  \bibinfo{author}{\bibfnamefont{N.}~\bibnamefont{{Aghanim}}},
  \bibinfo{author}{\bibfnamefont{M.}~\bibnamefont{{Arnaud}}},
  \bibinfo{author}{\bibfnamefont{F.}~\bibnamefont{{Arroja}}},
  \bibinfo{author}{\bibfnamefont{M.}~\bibnamefont{{Ashdown}}},
  \bibinfo{author}{\bibfnamefont{J.}~\bibnamefont{{Aumont}}},
  \bibinfo{author}{\bibfnamefont{C.}~\bibnamefont{{Baccigalupi}}},
  \bibinfo{author}{\bibfnamefont{M.}~\bibnamefont{{Ballardini}}},
  \bibinfo{author}{\bibfnamefont{A.~J.} \bibnamefont{{Banday}}},
  \bibnamefont{et~al.}, \bibinfo{journal}{ArXiv e-prints}
  (\bibinfo{year}{2015}{\natexlab{b}}), \eprint{1502.02114}.

\bibitem[{\citenamefont{{Meerburg} et~al.}(2012)\citenamefont{{Meerburg},
  {Wijers}, and {van der Schaar}}}]{PSOscillations2011}
\bibinfo{author}{\bibfnamefont{P.~D.} \bibnamefont{{Meerburg}}},
  \bibinfo{author}{\bibfnamefont{R.~A.~M.~J.} \bibnamefont{{Wijers}}},
  \bibnamefont{and} \bibinfo{author}{\bibfnamefont{J.~P.} \bibnamefont{{van der
  Schaar}}}, \bibinfo{journal}{\mnras} \textbf{\bibinfo{volume}{421}},
  \bibinfo{pages}{369} (\bibinfo{year}{2012}), \eprint{1109.5264}.

\bibitem[{\citenamefont{{Meerburg}
  et~al.}(2014{\natexlab{a}})\citenamefont{{Meerburg}, {Spergel}, and
  {Wandelt}}}]{MeerburgOscillations2014}
\bibinfo{author}{\bibfnamefont{P.~D.} \bibnamefont{{Meerburg}}},
  \bibinfo{author}{\bibfnamefont{D.~N.} \bibnamefont{{Spergel}}},
  \bibnamefont{and} \bibinfo{author}{\bibfnamefont{B.~D.}
  \bibnamefont{{Wandelt}}}, \bibinfo{journal}{ArXiv e-prints}
  (\bibinfo{year}{2014}{\natexlab{a}}), \eprint{1406.0548}.

\bibitem[{\citenamefont{{Meerburg}
  et~al.}(2014{\natexlab{b}})\citenamefont{{Meerburg}, {Spergel}, and
  {Wandelt}}}]{Meerburg2014a}
\bibinfo{author}{\bibfnamefont{P.~D.} \bibnamefont{{Meerburg}}},
  \bibinfo{author}{\bibfnamefont{D.~N.} \bibnamefont{{Spergel}}},
  \bibnamefont{and} \bibinfo{author}{\bibfnamefont{B.~D.}
  \bibnamefont{{Wandelt}}}, \bibinfo{journal}{\prd}
  \textbf{\bibinfo{volume}{89}}, \bibinfo{eid}{063537}
  (\bibinfo{year}{2014}{\natexlab{b}}), \eprint{1308.3705}.

\bibitem[{\citenamefont{{Meerburg}
  et~al.}(2014{\natexlab{c}})\citenamefont{{Meerburg}, {Spergel}, and
  {Wandelt}}}]{Meerburg2014b}
\bibinfo{author}{\bibfnamefont{P.~D.} \bibnamefont{{Meerburg}}},
  \bibinfo{author}{\bibfnamefont{D.~N.} \bibnamefont{{Spergel}}},
  \bibnamefont{and} \bibinfo{author}{\bibfnamefont{B.~D.}
  \bibnamefont{{Wandelt}}}, \bibinfo{journal}{\prd}
  \textbf{\bibinfo{volume}{89}}, \bibinfo{eid}{063536}
  (\bibinfo{year}{2014}{\natexlab{c}}), \eprint{1308.3704}.

\bibitem[{\citenamefont{{Easther} and
  {Flauger}}(2013)}]{PSOscillationsFlauger2013}
\bibinfo{author}{\bibfnamefont{R.}~\bibnamefont{{Easther}}} \bibnamefont{and}
  \bibinfo{author}{\bibfnamefont{R.}~\bibnamefont{{Flauger}}},
  \bibinfo{journal}{ArXiv e-prints}  (\bibinfo{year}{2013}),
  \eprint{1308.3736}.

\bibitem[{\citenamefont{{Flauger} et~al.}(2014)\citenamefont{{Flauger},
  {McAllister}, {Silverstein}, and {Westphal}}}]{DriftingOscillations2014}
\bibinfo{author}{\bibfnamefont{R.}~\bibnamefont{{Flauger}}},
  \bibinfo{author}{\bibfnamefont{L.}~\bibnamefont{{McAllister}}},
  \bibinfo{author}{\bibfnamefont{E.}~\bibnamefont{{Silverstein}}},
  \bibnamefont{and}
  \bibinfo{author}{\bibfnamefont{A.}~\bibnamefont{{Westphal}}},
  \bibinfo{journal}{ArXiv e-prints}  (\bibinfo{year}{2014}),
  \eprint{1412.1814}.

\bibitem[{\citenamefont{{Chen} et~al.}(2007)\citenamefont{{Chen}, {Easther},
  and {Lim}}}]{NGFeaturesChen2007}
\bibinfo{author}{\bibfnamefont{X.}~\bibnamefont{{Chen}}},
  \bibinfo{author}{\bibfnamefont{R.}~\bibnamefont{{Easther}}},
  \bibnamefont{and} \bibinfo{author}{\bibfnamefont{E.~A.} \bibnamefont{{Lim}}},
  \bibinfo{journal}{\jcap} \textbf{\bibinfo{volume}{6}}, \bibinfo{eid}{023}
  (\bibinfo{year}{2007}), \eprint{arXiv:astro-ph/0611645}.

\bibitem[{\citenamefont{{Holman} and
  {Tolley}}(2008)}]{InitialStateOriginalHolman2007}
\bibinfo{author}{\bibfnamefont{R.}~\bibnamefont{{Holman}}} \bibnamefont{and}
  \bibinfo{author}{\bibfnamefont{A.~J.} \bibnamefont{{Tolley}}},
  \bibinfo{journal}{\jcap} \textbf{\bibinfo{volume}{5}}, \bibinfo{eid}{001}
  (\bibinfo{year}{2008}), \eprint{0710.1302}.

\bibitem[{\citenamefont{{Meerburg} et~al.}(2009)\citenamefont{{Meerburg}, {van
  der Schaar}, and {Corasaniti}}}]{NonBDBispectrum2009}
\bibinfo{author}{\bibfnamefont{P.~D.} \bibnamefont{{Meerburg}}},
  \bibinfo{author}{\bibfnamefont{J.~P.} \bibnamefont{{van der Schaar}}},
  \bibnamefont{and} \bibinfo{author}{\bibfnamefont{P.~S.}
  \bibnamefont{{Corasaniti}}}, \bibinfo{journal}{\jcap}
  \textbf{\bibinfo{volume}{5}}, \bibinfo{eid}{018} (\bibinfo{year}{2009}),
  \eprint{0901.4044}.

\bibitem[{\citenamefont{{Meerburg} et~al.}(2010)\citenamefont{{Meerburg}, {van
  der Schaar}, and {Jackson}}}]{NonBDBispectrum2010}
\bibinfo{author}{\bibfnamefont{P.~D.} \bibnamefont{{Meerburg}}},
  \bibinfo{author}{\bibfnamefont{J.~P.} \bibnamefont{{van der Schaar}}},
  \bibnamefont{and} \bibinfo{author}{\bibfnamefont{M.~G.}
  \bibnamefont{{Jackson}}}, \bibinfo{journal}{\jcap}
  \textbf{\bibinfo{volume}{2}}, \bibinfo{eid}{001} (\bibinfo{year}{2010}),
  \eprint{0910.4986}.

\bibitem[{\citenamefont{{Chen}}(2010)}]{ResonantAndNonBDChen2010}
\bibinfo{author}{\bibfnamefont{X.}~\bibnamefont{{Chen}}},
  \bibinfo{journal}{\jcap} \textbf{\bibinfo{volume}{12}}, \bibinfo{eid}{003}
  (\bibinfo{year}{2010}), \eprint{1008.2485}.

\bibitem[{\citenamefont{{Meerburg} and {van der
  Schaar}}(2011)}]{NonBDBispectrum2010b}
\bibinfo{author}{\bibfnamefont{P.~D.} \bibnamefont{{Meerburg}}}
  \bibnamefont{and} \bibinfo{author}{\bibfnamefont{J.~P.} \bibnamefont{{van der
  Schaar}}}, \bibinfo{journal}{\prd} \textbf{\bibinfo{volume}{83}},
  \bibinfo{eid}{043520} (\bibinfo{year}{2011}), \eprint{1009.5660}.

\bibitem[{\citenamefont{{Agullo} et~al.}(2012)\citenamefont{{Agullo},
  {Navarro-Salas}, and {Parker}}}]{nonBDbospectrumPAgullo2011}
\bibinfo{author}{\bibfnamefont{I.}~\bibnamefont{{Agullo}}},
  \bibinfo{author}{\bibfnamefont{J.}~\bibnamefont{{Navarro-Salas}}},
  \bibnamefont{and} \bibinfo{author}{\bibfnamefont{L.}~\bibnamefont{{Parker}}},
  \bibinfo{journal}{\jcap} \textbf{\bibinfo{volume}{5}}, \bibinfo{eid}{019}
  (\bibinfo{year}{2012}), \eprint{1112.1581}.

\bibitem[{\citenamefont{{Agullo} and {Parker}}(2011)}]{MixedStateAgullo2011}
\bibinfo{author}{\bibfnamefont{I.}~\bibnamefont{{Agullo}}} \bibnamefont{and}
  \bibinfo{author}{\bibfnamefont{L.}~\bibnamefont{{Parker}}},
  \bibinfo{journal}{\prd} \textbf{\bibinfo{volume}{83}}, \bibinfo{eid}{063526}
  (\bibinfo{year}{2011}), \eprint{1010.5766}.

\bibitem[{\citenamefont{{Behbahani} et~al.}(2014)\citenamefont{{Behbahani},
  {Mirbabayi}, {Senatore}, and {Smith}}}]{NGsEFT2014}
\bibinfo{author}{\bibfnamefont{S.~R.} \bibnamefont{{Behbahani}}},
  \bibinfo{author}{\bibfnamefont{M.}~\bibnamefont{{Mirbabayi}}},
  \bibinfo{author}{\bibfnamefont{L.}~\bibnamefont{{Senatore}}},
  \bibnamefont{and} \bibinfo{author}{\bibfnamefont{K.~M.}
  \bibnamefont{{Smith}}}, \bibinfo{journal}{\jcap}
  \textbf{\bibinfo{volume}{11}}, \bibinfo{eid}{035} (\bibinfo{year}{2014}),
  \eprint{1407.7042}.

\bibitem[{\citenamefont{{Albrecht} et~al.}(2014)\citenamefont{{Albrecht},
  {Bolis}, and {Holman}}}]{Multiverse2014_Albrecht}
\bibinfo{author}{\bibfnamefont{A.}~\bibnamefont{{Albrecht}}},
  \bibinfo{author}{\bibfnamefont{N.}~\bibnamefont{{Bolis}}}, \bibnamefont{and}
  \bibinfo{author}{\bibfnamefont{R.}~\bibnamefont{{Holman}}},
  \bibinfo{journal}{Journal of High Energy Physics}
  \textbf{\bibinfo{volume}{11}}, \bibinfo{pages}{93} (\bibinfo{year}{2014}),
  \eprint{1408.6859}.

\bibitem[{\citenamefont{{Dimitrakopoulos}
  et~al.}(2015)\citenamefont{{Dimitrakopoulos}, {Kabir}, {Mosk}, {Parikh}, and
  {van der Schaar}}}]{JP2015}
\bibinfo{author}{\bibfnamefont{F.~V.} \bibnamefont{{Dimitrakopoulos}}},
  \bibinfo{author}{\bibfnamefont{L.}~\bibnamefont{{Kabir}}},
  \bibinfo{author}{\bibfnamefont{B.}~\bibnamefont{{Mosk}}},
  \bibinfo{author}{\bibfnamefont{M.}~\bibnamefont{{Parikh}}}, \bibnamefont{and}
  \bibinfo{author}{\bibfnamefont{J.~P.} \bibnamefont{{van der Schaar}}},
  \bibinfo{journal}{ArXiv e-prints}  (\bibinfo{year}{2015}),
  \eprint{1502.00113}.

\bibitem[{\citenamefont{{M{\"u}nchmeyer}
  et~al.}(2014)\citenamefont{{M{\"u}nchmeyer}, {Bouchet}, {Jackson}, and
  {Wandelt}}}]{LinearOscillationsMoritz2014}
\bibinfo{author}{\bibfnamefont{M.}~\bibnamefont{{M{\"u}nchmeyer}}},
  \bibinfo{author}{\bibfnamefont{F.}~\bibnamefont{{Bouchet}}},
  \bibinfo{author}{\bibfnamefont{M.~G.} \bibnamefont{{Jackson}}},
  \bibnamefont{and}
  \bibinfo{author}{\bibfnamefont{B.}~\bibnamefont{{Wandelt}}},
  \bibinfo{journal}{\aap} \textbf{\bibinfo{volume}{570}}, \bibinfo{eid}{A94}
  (\bibinfo{year}{2014}), \eprint{1405.2550}.

\bibitem[{\citenamefont{Mukhanov}(1988)}]{Mukhanov:1988jd}
\bibinfo{author}{\bibfnamefont{V.~F.} \bibnamefont{Mukhanov}},
  \bibinfo{journal}{Sov.Phys.JETP} \textbf{\bibinfo{volume}{67}},
  \bibinfo{pages}{1297} (\bibinfo{year}{1988}).

\bibitem[{\citenamefont{{Mukhanov} et~al.}(1992)\citenamefont{{Mukhanov},
  {Feldman}, and {Brandenberger}}}]{1992PhR...215..203M}
\bibinfo{author}{\bibfnamefont{V.~F.} \bibnamefont{{Mukhanov}}},
  \bibinfo{author}{\bibfnamefont{H.~A.} \bibnamefont{{Feldman}}},
  \bibnamefont{and} \bibinfo{author}{\bibfnamefont{R.~H.}
  \bibnamefont{{Brandenberger}}}, \bibinfo{journal}{\physrep}
  \textbf{\bibinfo{volume}{215}}, \bibinfo{pages}{203} (\bibinfo{year}{1992}).

\bibitem[{\citenamefont{{Martin} and
  {Brandenberger}}(2001)}]{TransPlanckianMartin2001}
\bibinfo{author}{\bibfnamefont{J.}~\bibnamefont{{Martin}}} \bibnamefont{and}
  \bibinfo{author}{\bibfnamefont{R.~H.} \bibnamefont{{Brandenberger}}},
  \bibinfo{journal}{\prd} \textbf{\bibinfo{volume}{63}}, \bibinfo{eid}{123501}
  (\bibinfo{year}{2001}), \eprint{hep-th/0005209}.

\bibitem[{\citenamefont{{Danielsson}}(2002)}]{DanielssonTP2002}
\bibinfo{author}{\bibfnamefont{U.~H.} \bibnamefont{{Danielsson}}},
  \bibinfo{journal}{\prd} \textbf{\bibinfo{volume}{66}}, \bibinfo{eid}{023511}
  (\bibinfo{year}{2002}), \eprint{hep-th/0203198}.

\bibitem[{\citenamefont{Martin and Brandenberger}(2003)}]{Martin:2003kp}
\bibinfo{author}{\bibfnamefont{J.}~\bibnamefont{Martin}} \bibnamefont{and}
  \bibinfo{author}{\bibfnamefont{R.}~\bibnamefont{Brandenberger}},
  \bibinfo{journal}{Phys.Rev.} \textbf{\bibinfo{volume}{D68}},
  \bibinfo{pages}{063513} (\bibinfo{year}{2003}), \eprint{hep-th/0305161}.

\bibitem[{\citenamefont{{Martin} and
  {Ringeval}}(2004)}]{PSOscillationsMartin2004}
\bibinfo{author}{\bibfnamefont{J.}~\bibnamefont{{Martin}}} \bibnamefont{and}
  \bibinfo{author}{\bibfnamefont{C.}~\bibnamefont{{Ringeval}}},
  \bibinfo{journal}{\prd} \textbf{\bibinfo{volume}{69}}, \bibinfo{eid}{083515}
  (\bibinfo{year}{2004}), \eprint{arXiv:astro-ph/0310382}.

\bibitem[{\citenamefont{{Greene}
  et~al.}(2005{\natexlab{a}})\citenamefont{{Greene}, {Schalm}, {van der
  Schaar}, and {Shiu}}}]{GreeneEtAl2005}
\bibinfo{author}{\bibfnamefont{B.}~\bibnamefont{{Greene}}},
  \bibinfo{author}{\bibfnamefont{K.}~\bibnamefont{{Schalm}}},
  \bibinfo{author}{\bibfnamefont{J.~P.} \bibnamefont{{van der Schaar}}},
  \bibnamefont{and} \bibinfo{author}{\bibfnamefont{G.}~\bibnamefont{{Shiu}}},
  in \emph{\bibinfo{booktitle}{22nd Texas Symposium on Relativistic
  Astrophysics}}, edited by
  \bibinfo{editor}{\bibfnamefont{P.}~\bibnamefont{{Chen}}},
  \bibinfo{editor}{\bibfnamefont{E.}~\bibnamefont{{Bloom}}},
  \bibinfo{editor}{\bibfnamefont{G.}~\bibnamefont{{Madejski}}},
  \bibnamefont{and}
  \bibinfo{editor}{\bibfnamefont{V.}~\bibnamefont{{Patrosian}}}
  (\bibinfo{year}{2005}{\natexlab{a}}), pp. \bibinfo{pages}{1--8},
  \eprint{arXiv:astro-ph/0503458}.

\bibitem[{\citenamefont{{Greene}
  et~al.}(2005{\natexlab{b}})\citenamefont{{Greene}, {Schalm}, {Shiu}, and {van
  der Schaar}}}]{BackreactionInitialState2005}
\bibinfo{author}{\bibfnamefont{B.~R.} \bibnamefont{{Greene}}},
  \bibinfo{author}{\bibfnamefont{K.}~\bibnamefont{{Schalm}}},
  \bibinfo{author}{\bibfnamefont{G.}~\bibnamefont{{Shiu}}}, \bibnamefont{and}
  \bibinfo{author}{\bibfnamefont{J.~P.} \bibnamefont{{van der Schaar}}},
  \bibinfo{journal}{\jcap} \textbf{\bibinfo{volume}{2}}, \bibinfo{eid}{001}
  (\bibinfo{year}{2005}{\natexlab{b}}), \eprint{hep-th/0411217}.

\bibitem[{\citenamefont{{Brandenberger} and
  {Martin}}(2013)}]{2013CQGra..30k3001B}
\bibinfo{author}{\bibfnamefont{R.~H.} \bibnamefont{{Brandenberger}}}
  \bibnamefont{and} \bibinfo{author}{\bibfnamefont{J.}~\bibnamefont{{Martin}}},
  \bibinfo{journal}{Classical and Quantum Gravity}
  \textbf{\bibinfo{volume}{30}}, \bibinfo{eid}{113001} (\bibinfo{year}{2013}),
  \eprint{1211.6753}.

\bibitem[{\citenamefont{{Schalm} et~al.}(2004)\citenamefont{{Schalm}, {Shiu},
  and {van der Schaar}}}]{BEFTInitialState2005}
\bibinfo{author}{\bibfnamefont{K.}~\bibnamefont{{Schalm}}},
  \bibinfo{author}{\bibfnamefont{G.}~\bibnamefont{{Shiu}}}, \bibnamefont{and}
  \bibinfo{author}{\bibfnamefont{J.~P.} \bibnamefont{{van der Schaar}}},
  \bibinfo{journal}{Journal of High Energy Physics}
  \textbf{\bibinfo{volume}{4}}, \bibinfo{eid}{076} (\bibinfo{year}{2004}),
  \eprint{hep-th/0401164}.

\bibitem[{\citenamefont{{Flauger} et~al.}(2010)\citenamefont{{Flauger},
  {McAllister}, {Pajer}, {Westphal}, and {Xu}}}]{MonodromyFlauger2009}
\bibinfo{author}{\bibfnamefont{R.}~\bibnamefont{{Flauger}}},
  \bibinfo{author}{\bibfnamefont{L.}~\bibnamefont{{McAllister}}},
  \bibinfo{author}{\bibfnamefont{E.}~\bibnamefont{{Pajer}}},
  \bibinfo{author}{\bibfnamefont{A.}~\bibnamefont{{Westphal}}},
  \bibnamefont{and} \bibinfo{author}{\bibfnamefont{G.}~\bibnamefont{{Xu}}},
  \bibinfo{journal}{\jcap} \textbf{\bibinfo{volume}{6}}, \bibinfo{eid}{009}
  (\bibinfo{year}{2010}), \eprint{0907.2916}.

\bibitem[{\citenamefont{Maldacena}(2003)}]{Maldacena:2002vr}
\bibinfo{author}{\bibfnamefont{J.~M.} \bibnamefont{Maldacena}},
  \bibinfo{journal}{JHEP} \textbf{\bibinfo{volume}{0305}}, \bibinfo{pages}{013}
  (\bibinfo{year}{2003}), \eprint{astro-ph/0210603}.

\bibitem[{\citenamefont{{Porrati}}(2004)}]{BoundaryNGsPorrati2004}
\bibinfo{author}{\bibfnamefont{M.}~\bibnamefont{{Porrati}}},
  \bibinfo{journal}{ArXiv High Energy Physics - Theory e-prints}
  (\bibinfo{year}{2004}), \eprint{hep-th/0409210}.

\bibitem[{\citenamefont{{Flauger} et~al.}(2013)\citenamefont{{Flauger},
  {Green}, and {Porto}}}]{ExcitedLimitBispectra}
\bibinfo{author}{\bibfnamefont{R.}~\bibnamefont{{Flauger}}},
  \bibinfo{author}{\bibfnamefont{D.}~\bibnamefont{{Green}}}, \bibnamefont{and}
  \bibinfo{author}{\bibfnamefont{R.~A.} \bibnamefont{{Porto}}},
  \bibinfo{journal}{\jcap} \textbf{\bibinfo{volume}{8}}, \bibinfo{eid}{032}
  (\bibinfo{year}{2013}), \eprint{1303.1430}.

\bibitem[{\citenamefont{{Babich} et~al.}(2004)\citenamefont{{Babich},
  {Creminelli}, and {Zaldarriaga}}}]{NonGaussianShapes}
\bibinfo{author}{\bibfnamefont{D.}~\bibnamefont{{Babich}}},
  \bibinfo{author}{\bibfnamefont{P.}~\bibnamefont{{Creminelli}}},
  \bibnamefont{and}
  \bibinfo{author}{\bibfnamefont{M.}~\bibnamefont{{Zaldarriaga}}},
  \bibinfo{journal}{\jcap} \textbf{\bibinfo{volume}{8}}, \bibinfo{eid}{009}
  (\bibinfo{year}{2004}), \eprint{astro-ph/0405356}.

\bibitem[{\citenamefont{{Agarwal} et~al.}(2013)\citenamefont{{Agarwal},
  {Holman}, {Tolley}, and {Lin}}}]{EFTexcitedStates}
\bibinfo{author}{\bibfnamefont{N.}~\bibnamefont{{Agarwal}}},
  \bibinfo{author}{\bibfnamefont{R.}~\bibnamefont{{Holman}}},
  \bibinfo{author}{\bibfnamefont{A.~J.} \bibnamefont{{Tolley}}},
  \bibnamefont{and} \bibinfo{author}{\bibfnamefont{J.}~\bibnamefont{{Lin}}},
  \bibinfo{journal}{Journal of High Energy Physics}
  \textbf{\bibinfo{volume}{5}}, \bibinfo{eid}{85} (\bibinfo{year}{2013}),
  \eprint{1212.1172}.

\bibitem[{\citenamefont{{Wagner} et~al.}(2010)\citenamefont{{Wagner}, {Verde},
  and {Boubekeur}}}]{BiasVerde2010}
\bibinfo{author}{\bibfnamefont{C.}~\bibnamefont{{Wagner}}},
  \bibinfo{author}{\bibfnamefont{L.}~\bibnamefont{{Verde}}}, \bibnamefont{and}
  \bibinfo{author}{\bibfnamefont{L.}~\bibnamefont{{Boubekeur}}},
  \bibinfo{journal}{\jcap} \textbf{\bibinfo{volume}{10}}, \bibinfo{eid}{022}
  (\bibinfo{year}{2010}), \eprint{1006.5793}.

\end{thebibliography}

\end{document}